\documentclass[12pt, draftclsnofoot, onecolumn]{IEEEtran}
\IEEEoverridecommandlockouts                       
\usepackage{cite}
\usepackage{amsmath,amssymb,amsfonts}
\usepackage{algorithmic}
\usepackage{graphicx}
\usepackage{paralist}
\usepackage{textcomp}
\usepackage{mathtools}
\usepackage{xcolor}
\usepackage{color}
\usepackage{tcolorbox}
\usepackage{multirow}
\usepackage{lipsum}
\usepackage{verbatim}
\usepackage[normalem]{ulem}
\usepackage{dsfont}
\usepackage{listings}
\usepackage{url}
\usepackage[utf8]{inputenc}
\graphicspath{ {./images/} }
\usepackage{acronym}
\usepackage{cite}
\usepackage{lettrine}
\usepackage{subfigure}
\usepackage{enumitem}
\usepackage{array}
\usepackage{amsthm}
\newcommand*{\Comb}[2]{{}^{#1}C_{#2}}%
\newcommand{\RomanNumeralCaps}[1]
    {\MakeUppercase{\romannumeral #1}}

\newtheorem{corollary}{Corollary}[]
\newtheorem{lemma}{Lemma}[]
\newtheorem{proposition}{Proposition}[]
\def\BibTeX{{\rm B\kern-.05em{\sc i\kern-.025em b}\kern-.08em
    T\kern-.1667em\lower.7ex\hbox{E}\kern-.125emX}}
    \newcolumntype{M}[1]{>{\centering\arraybackslash}m{#1}}
\begin{document}
\acrodef{PPP}[PPP]{poisson point process} 
\acrodef{3D}[3D]{3-dimension} 
\acrodef{BS}[BS]{base station} 
\acrodef{AP}[AP]{access point}
\acrodef{TBS}[TBS]{terrestrial base station} 
\acrodef{UAV}[UAV]{unmanned aerial vehicle}
\acrodef{BS}[BS]{base station}
\acrodef{MPC}[MPC]{most popular content}
\acrodef{LPC}[LPC]{less popular content}
\acrodef{UAV-BS}[UAV-BS]{UAV-Base station}
\acrodef{UAV-AP}[UAV-AP]{UAV-Access point}
\acrodef{LoS}[LoS]{line-of-sight}
\acrodef{MBS}[MBS]{macro base station}
\acrodef{SBS}[SBS]{small base station}
\acrodef{IAX}[IAX]{integrated access and x- haul}
\acrodef{IAB}[IAB]{integrated access and backhaul}
\acrodef{NLoS}[NLoS]{Non-line-of-sight}
\acrodef{SINR}[SINR]{signal to interference noise ratio}
\acrodef{HetNet}[HetNet]{heterogeneous network}
\acrodef{QoS}[QoS]{quality of service}
\title{\fontsize{22}{22}\selectfont Cache Enabled UAV HetNets: Access - xHaul Coverage Analysis and Optimal Resource Partitioning}
\vspace{-10cm}
\author{Neetu R.R, Gourab Ghatak, Anand Srivastava and Vivek Ashok Bohara
\thanks{The authors are with the Department of Electronics and Communication Engineering, IIIT-Delhi, India. email: \{neetur, gourab.ghatak, anand, vivek.b\} @iiitd.ac.in.}}

\maketitle
\vspace{-2cm}
\begin{abstract}
We study an urban wireless network in which cache-enabled \acp{UAV-AP} and \acp{UAV-BS} are deployed to provide higher throughput and ad-hoc coverage to users on the ground. The cache-enabled \acp{UAV-AP} route the user data to the core network via either \acp{TBS} or backhaul-enabled \acp{UAV-BS} through an xHaul link. 
First, we derive the association probabilities in the access and xHaul links. Interestingly, we show that to maximize the \ac{LoS} \ac{UAV} association, densifying the \ac{UAV} deployment may not be beneficial after a threshold. Then, we obtain the \ac{SINR} coverage probability of the typical user in the access link and the tagged \ac{UAV-AP} in the xHaul link, respectively. The \ac{SINR} coverage analysis is employed to characterize the successful content delivery probability by jointly considering the probability of successful access and xHaul transmissions and successful cache-hit probability. \textcolor{blue}{We numerically optimize the distribution of frequency resources between the access and the xHaul links to maximize the successful content delivery to the users.} For a given storage capacity at the \acp{UAV}, our study prescribes the network operator optimal bandwidth partitioning factors and dimensioning rules concerning the deployment of the \acp{UAV-AP}.
\end{abstract}
\vspace{-0.6cm}
\begin{IEEEkeywords}
\vspace{-0.3cm}
Cache-enabled \acp{UAV}, optimal resource allocation, success probability, 3D placement, xHaul.
\end{IEEEkeywords}
\vspace{-0.5cm}
\section{Introduction}
\lettrine{U}{nmanned} aerial vehicles (UAVs), mounted with remote radio heads (RRHs) can act as access points (either aerial relays or \acp{BS}) to deliver reliable, cost-effective, and on-demand wireless connectivity to the ground users. They potentially enhance the coverage and the capacity of cellular and ad-hoc networks. Such \ac{UAV} networks have found applications in disaster relief scenarios~\cite{16}, wireless sensor networks~\cite{17} and capacity augmentation in high-traffic areas~\cite{18}. The ability of the \acp{UAV-AP} to adjust their 3D position in real-time can facilitate unobstructed line-of-sight (LoS) links to the ground users and maintain a reliable connection to the terrestrial base stations (TBSs) for the transport of the user data to the core network. Additionally, provisioning local storage at the \acp{UAV-AP} by proactive caching can reduce the latency of the user applications while simultaneously reducing the load from the backhaul network.
One of the most significant issues in the deployment planning of \ac{UAV}-aided wireless networks is the design of backhaul links and its optimization with respect to user throughput by jointly taking into account the user density and the cache size. Based on the functionality available at the \ac{UAV-AP}, specifically, the centralized and distributed units are split in the architecture~\cite{35}, the link from the \ac{UAV-AP} to a \ac{UAV-BS} or to a \ac{TBS} can be classified as either a fronthaul or a midhaul. In this paper, we use the term {\it xHaul} to denote either of these types of links. The optimization of the access and xHaul is particularly challenging due to the temporally varying UAV and, user locations as well as the user requirements. In this regard, stochastic geometry provides an efficient tool to characterize the performance of \ac{UAV} networks by assuming the locations of the \acp{UAV} and users as a spatial stochastic process and evaluating the key performance indicators (KPIs) in an expected sense. This equips the operator with essential dimensioning and initial deployment insights for such networks. Consequently, in the proposed work, we develop a stochastic geometry model to jointly study the \ac{UAV} access and xHaul links. Also, we investigate the optimal distribution of frequency resources between the access and xHaul network to maximize the content delivery success probability at the users by taking into account the storage capacity of the local cache at the \acp{UAV-AP}.
\vspace{-0.6cm}
\subsection{Related Work}
There has been an increasing interest in the use of \acp{UAV} as aerial \acp{BS} or relays, e.g., \cite{21} -\nocite{22} \cite{23}. Integrating \acp{UAV} with legacy cellular networks results in vertical heterogeneous networks (HetNets)~\cite{hetnets}, which offer increased flexibility to the operator and enhanced coverage of the users. Additionally, provisioning of storage at the UAVs further augment the data rate and reduce the latency of services~\cite{cache}. A comprehensive survey on UAV-assisted cellular communications can be found in the reference~\cite{24}. Contrary to \acp{TBS}, \acp{UAV} due to their controllable altitudes, can facilitate a higher probability of a direct \ac{LoS} link to the users. In order to characterize the visibility conditions, authors in~\cite{25} have presented a tractable model that takes into account the height of buildings, the ratio of built-up area to total land area, and the number of buildings per unit area. The authors have prescribed various visibility scenarios like suburban, urban, dense urban, and highrise urban.
It may be noted that although the access link, i.e., UAV to user link, can be in a strong \ac{LoS} state, the backhaul link, i.e., the UAV to the core network link, can potentially cause a bottleneck for high throughput or reliability-constrained applications~\cite{32}. In this regard, \ac{IAB} is an attractive technology that efficiently exploits the same bandwidth to facilitate both access and backhaul connectivity~\cite{33}. The potentials and challenges of \ac{IAB} for 5G mm-wave networks were investigated in~\cite{26}. The authors have highlighted the augmented throughput offered by \ac{IAB} and the reduction in deployment costs. In our work, we investigate a scheme that partitions the bandwidth between the access and the backhaul links and optimize it for user throughput. Such a trade-off factor for resource allocation was studied in~\cite{8}. Furthermore, based on the achievable rate, they have formulated a resource allocation problem to enhance the user's data rate. However, they have not studied the impact of caching and the probability of successful content delivery at the user end. Additionally, from their work, the investigation of the impact of the intensity of \acp{UAV} on the association of user to the \acp{BS} for various visibility scenarios is missing, which can only be carried out either by extensive system-level simulations or using spatial stochastic processes.

The backhaul capacity impacts the placement of UAVs as well, which was investigated in~\cite{27}, where the authors have proposed a backhaul-limited optimal \ac{UAV-BS} placement algorithm and studied the effect of the user mobility on the \ac{UAV} placement. Furthermore, the authors in~\cite{28} have investigated the trends in user-\ac{BS} association with an increasing number of users for various caching schemes and corresponding bandwidth allocations while minimizing the total downlink transmit power. They also discussed the total backhaul capacity usage in aerial \acp{BS} while increasing the number of users in access links for different caching strategies. However, ~\cite{28} does not explore the trade-offs in frequency resource partitioning between access and backhaul links. Specifically in backhaul-constrained networks, proactive caching can improve the system performance by reducing the backhaul load. In~\cite{7}, the authors have proposed a caching scheme by managing the content popularity to improve the success probability. They have analyzed the impact of the density of \acp{UAV-BS}, caching capacity, and the altitude of the \acp{UAV} on the successful content delivery, energy efficiency, and coverage probability of the network. The authors in~\cite{29} have formulated an optimization problem to minimize content delivery delay in \ac{UAV}-non-orthogonal multiple-access networks. They have developed a reinforcement learning algorithm to investigate the effect of cache capacity, the number of cache contents, and the number of users on the content delivery delay. In this line of work, the authors in~\cite{30} have investigated the content distribution by offloading the traffic in hotspot areas by the combination of \acp{UAV} and edge caching. They have evaluated a mean opinion score (MOS) to obtain the quality of experience (QoE) of the users considering the user association, \ac{UAV} placement and caching placement. Then a joint optimization problem is developed to maximize the MOS. 
\vspace{-0.6cm}
\subsection{Motivation and Contribution}
\textcolor{blue}{In \cite{29} and \cite{30}, the authors have proposed efficient algorithms to optimize certain network parameters like Mean Opinion Score (MOS), coverage, content delivery delay for a given realization of the network. However, using stochastic geometry-based analysis, we have given an expected view of the network by spatially averaging across all such network realizations. For example, the distribution of the efficacy of the proposed algorithms in~\cite{29} and~\cite{30} is challenging to derive, which may be possible using a
stochastic geometry-based study. Also, } in most of the prior works, the authors assume that the altitude of all the \ac{UAV} access points to be the same and, consequently, optimize it with respect to the user metrics. 
Motivated by this, we study a 3D spatial stochastic process to model the location of the UAVs. This is particularly challenging due to the requirement of distance distributions of UAVs restricted on a 3D half-plane.  Additionally, the joint impact of caching and optimal distribution of frequency resources on user performance have not been studied in the existing research works. To investigate this, we propose a cache-enabled \ac{IAX} \ac{UAV} wireless network overlaid on top of a legacy \ac{TBS} network to sustain the QoS requirements of the ground users. The main contributions of this paper are summarized as follows:
\begin{enumerate}
     \item \textcolor{blue}{We derive the distance distribution of (i) the nearest point on a 2-D plane from a typical point on a 3-D half-plane and (ii) the nearest point on a 3-D half-plane from a tagged point on the same 3-D half-plane. Although the deployment of \acp{UAV} and their real-time locations span the 3D space, these spatial properties have previously not been reported in the literature on stochastic geometry-based models and are particularly challenging due to the non-isotropic nature of the spatial process in a 3D sense. }
    \item Leveraging these results,
    we derive the association probabilities of the typical user with the \ac{LoS}/\ac{NLoS} \acp{UAV-AP} and the \acp{TBS} for the access link. \textcolor{blue}{We explored the impact of densification of the network on the \ac{LoS} link association.} Furthermore, we derive the xHaul association probabilities of the tagged \ac{UAV-AP} with either the \acp{UAV-BS} tier or the \acp{TBS} tier. \textcolor{blue}{A major challenge in such a characterization is the dependence of the xHaul link association probabilities on the access link association events.} To the best of our knowledge, this paper is the first work that mathematically characterizes this dependence and derives complete access - xHaul association framework.
    \item Based on the derived association probabilities, we obtain analytical expressions for the \ac{SINR} coverage probability of the typical user associated to a \ac{LoS}/\ac{NLoS} \acp{UAV} or \acp{TBS} in the access link. \textcolor{blue}{Additionally, we derive the \ac{SINR} coverage probability of the tagged \ac{UAV-AP} associated to \ac{UAV-BS} or \ac{TBS} in the xHaul link by taking into account the statistical dependence of the access and the backhaul distances.} The derived analytical expressions are then verified with extensive Monte-Carlo simulations.
    \item In this network, we also study a caching scheme where the subset of most popular contents are always stored locally at the UAV, while the remaining files are probabilistically cached. \textcolor{blue}{ We analyze\sout{d} the impact of caching in the access-xHaul resource allocation. We optimize the resource partitioning factor between the access and xHaul for different cache sizes, the number of users, and the density of \acp{UAV}. We optimize the service success probability that jointly considers into account the end-to-end SINR coverage of the access link, the xHaul link, and the cache hit event.} This reveals several key system designs and initial deployment insights to the network operator for deploying \ac{UAV}-aided cellular networks.
\end{enumerate}
The rest of the paper is organized as follows. In Section~\ref{sec:SM} we introduce our network model and outline the study objectives. Section~\ref{sec:dist} derives the relevant distance distributions and association probabilities. Based on this, in Sections~\ref{sec:SINR} and Section~\ref{sec:rate}, we derive the \ac{SINR} coverage probability and the content delivery success probability, respectively. Then, in Section~\ref{sec:result} we validate our analytical framework and present some numerical results to discuss the salient features of the network. Finally, the paper concludes in Section~\ref{sec:Con}.
\vspace{-0.5cm}
\section{System Model}
\label{sec:SM}
\textcolor{blue}{We consider a downlink  heterogeneous network (HetNet) deployed for scenarios where the available wireless access infrastructure is insufficient, e.g., during mass events in areas not prepared for large crowds.}
\begin{figure}
\centering
\includegraphics[width=.5\linewidth,height=.25\linewidth]{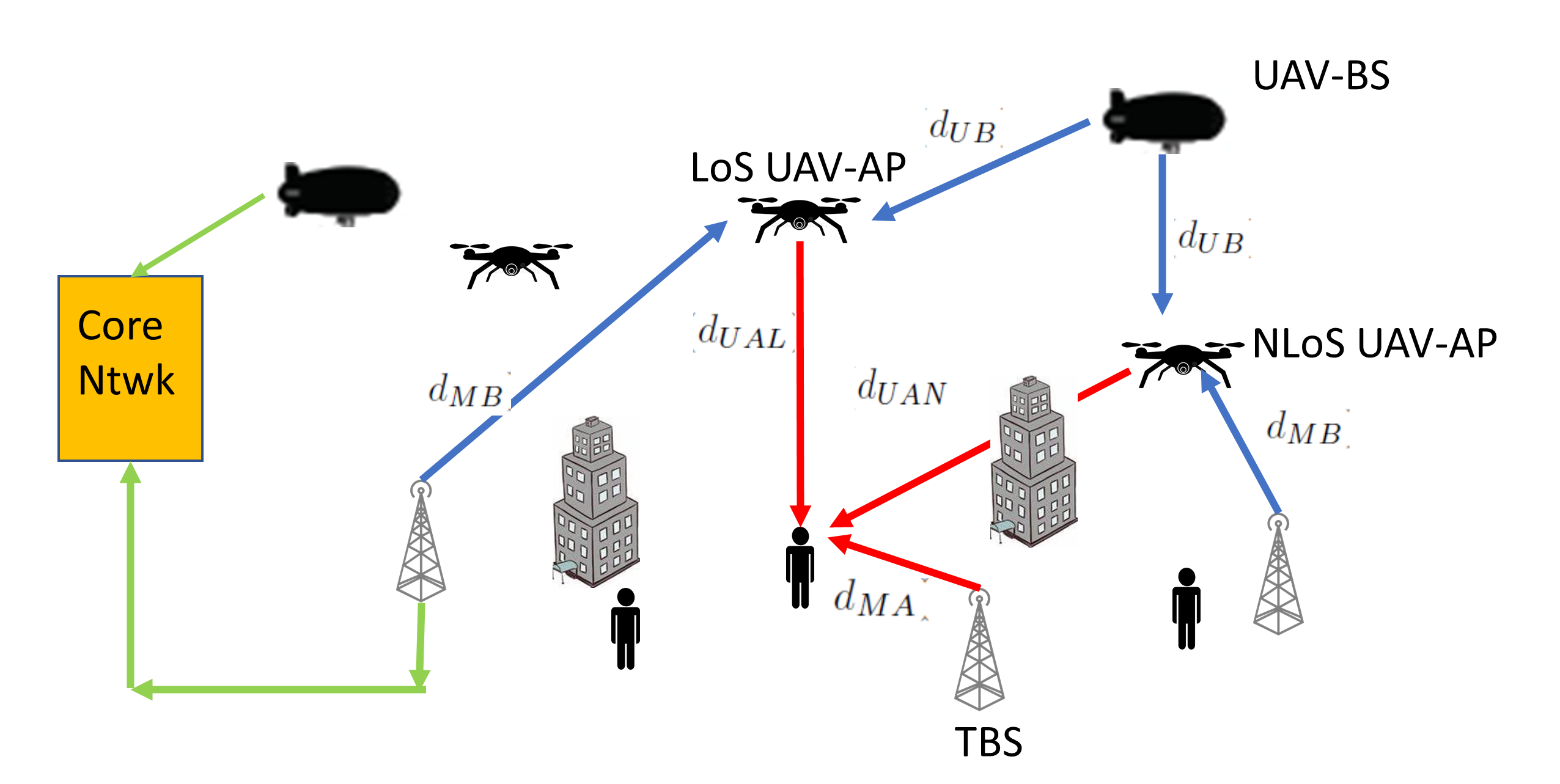}
\vspace{-1cm}
\caption{An illustration of the system model showing different distance variables considered. The access and the xHaul links are shown in red and blue, respectively. The backhaul link to the core network is shown in green is not discussed in this paper.}
\label{fig:system_main}
\end{figure}
 \textcolor{blue}{The network consists of cache-enabled \acp{UAV-AP} overlaid on top of a legacy \acp{TBS} consisting of \acp{MBS} and \acp{SBS}. The \acp{UAV-AP} are small-sized low platform aerial vehicles, which connect to either the \ac{TBS} or backhaul connected \ac{UAV-BS} for xHaul support as given in Fig.\ref{fig:system_main}. The \acp{UAV-BS} are large-sized aerial vehicles that are deployed at higher altitudes for a good xHaul connection with the core network.} The users are assumed to be located co-planar with the \ac{TBS}.  The locations of the cached-enabled \acp{UAV-AP} are modeled as a 3D \ac{PPP} $\Phi_U$ defined on $\mathbb{R}^2 \times \mathbb{R}^+$, with intensity $\lambda_{UA}$. On the contrary, the locations of the \acp{TBS} are modeled as a 2D PPP, $\Phi_M$ on $\mathbb{R}^2$, with intensity $\lambda_{M}$. \textcolor{blue}{Additionally, the locations of \acp{UAV-BS} are modeled as a 3D PPP $\Phi_B$ in $\mathbb{R}^2 \times \mathbb{R}^+$, with intensity $\lambda_{UB}$.} Further, it is assumed that $\Phi_B$ is independent of $\Phi_U$. Let $\Phi$=\{$X_l$\} be the point process which is the union of all independent \ac{PPP} in the network. Therefore, $\Phi$= $\Phi_M \cup \Phi_U \cup \Phi_B$. Without loss of generality, we perform the downlink analysis from the perspective of a typical user located at the origin of the 3D Euclidean space. The typical user associates to either the \ac{UAV-AP} tier or the \ac{TBS} tier, \textcolor{blue}{based on strongest BS association scheme in the access link. The received signal strength indicator (RSSI) measurements in the downlink access channel are estimated for the association of typical user to the \ac{UAV-AP} tier or the \ac{TBS} tier.  The association of the typical user to the \ac{UAV-BS} is limited because of its high altitude.} Given a \ac{UAV-AP} association, in case the file requested by the user is present in the \ac{UAV} cache, only the access link is used~\cite{10}. Otherwise, the \ac{UAV-AP} retrieves the file from either the \ac{TBS} tier or the \ac{UAV-BS} tier via an xHaul link. \textcolor{blue}{In this work, we do not consider any local storage at the \ac{UAV-BS} and the \ac{TBS} tier due to the assumption of a reliable backhaul connection to the core network from these tiers.} Thus, the \ac{UAV-AP} associates with either the \ac{TBS} tier or the \ac{UAV} \ac{BS} tier for xHaul transport, based on RSSI measurements. \textcolor{blue}{The \ac{UAV-BS} increases the xHaul capacity by enabling additional wireless xHaul links.} We partition the total available bandwidth $B$ between the access and the xHaul links using a bandwidth allocation factor, $\beta$ which can take any value between 0 and 1, $\beta\in [0, 1]$. The allocated bandwidth to the access link is $\beta B$, and that to the xHaul link is $(1-\beta)B$~\cite{8}. Moreover, we assume orthogonal frequency allocation to multiplex multiple users in the access link~\cite{13}. The downlink transmit powers of the \acp{UAV-AP}, the backhaul connected \ac{UAV} BSs, and the \acp{TBS} are $P_{UA}$, $P_{UB}$, and $P_{M}$, respectively. 
\vspace{-0.6cm}
\subsection{Channel Model: Access link}
The access link propagation consists of small-scale fading and large-scale path loss. Specifically, the \acp{TBS} transmissions experience small scale Rayleigh fading, $g_M$, with a variance of 1~\cite{31}. The \acp{UAV-AP} can either be in \ac{LoS} or \ac{NLoS} state from the perspective of the typical user. Let the locations of the \acp{UAV-AP} in LoS and NLoS be denoted as $\Phi_{L}$ and $\Phi_{N}$, respectively, where $\Phi_{U} = \Phi_{L} \cup \Phi_{N}$. The  probability of \ac{LoS} link between the \ac{UAV-AP} and the typical user is given as~\cite{3}:  
\vspace{-0.3cm}
\textcolor{blue}{\begin{equation}
       W_L(d,h)= \frac{1}{1+\eta\exp{\left(-\mu\left(\frac{180}{\pi}\sin^{-1}\left(\frac{h}{d}\right)-\eta\right)\right)}}
\label{eqn1}
\end{equation}}
\textcolor{blue}{where height $h$ of the \ac{UAV-AP} from the ground, which can be written in terms of the distance between the \ac{LoS} \ac{UAV-AP} and the typical user $d$, and $\theta$ which is the polar angle (angle between the polar axis and the line joining the typical user to the \ac{UAV-AP}).}
\begin{equation}
       W_L(\theta)= \frac{1}{1+\eta\exp{\left(-\mu\left(\frac{180}{\pi}\sin^{-1}\left(\frac{d cos(\theta)}{d}\right)-\eta\right)\right)}}
\label{eqn1}
\end{equation}

Thus, $d\cos$($\theta$) is the height of the \ac{UAV-AP} from the ground. From (\ref{eqn1}), it is evident that the probability of \ac{LoS} transmission is independent of the distance $d$ and only depends on $\theta$ and on the environment parameters. Here $\eta$ and $\mu$ are the environment parameters for different visibility scenarios like suburban, urban, dense urban and high-rise urban. The $\eta$ and $\mu$ values for different scenarios are suburban (4.88, 0.43), urban (9.61, 0.16), dense urban (11.95, 0.136), and high-rise urban (24.23, 0.08). Consequently, the probability of \ac{NLoS} transmissions are given as $W_N(\theta)= 1-W_L(\theta)$.  
Due to the higher local scattering, the propagation model from an \ac{NLoS} UAV to the typical user suffers from a  Rayleigh fading $g_{NL}$. On the contrary, for the \ac{LoS} UAV transmissions, we assume a Nakagami fading distribution, $G_L$, with shape parameter $m$~\cite{11}.
For the large scale path loss, we consider the classical power law where, the received power at the typical user from a \ac{TBS} at a distance of $d_{MA}$, an \ac{LoS} UAV at a distance of $d_{UAL}$, and an \ac{NLoS} UAV at a distance of $d_{UAN}$ is given by $R_{MA}= K_MP_{M}g_M(d_{MA})^{-\alpha_N}$, $R_{UAL}= K_UP_{UA}G_L(d_{UAL})^{-\alpha_L}$, and $R_{UAN}= K_UP_{UA}g_{NL}(d_{UAN})^{-\alpha_N}$, respectively. Here $K_U$ and $K_M$ are the path loss coefficients given by $K_U= K_M =(\frac{\lambda_c}{4\pi})^2$ where $\lambda_c$ is the carrier wavelength. Whereas, $\alpha_L$ and $\alpha_N$, are the path loss exponents.
\vspace{-0.2cm}
\renewcommand{\arraystretch}{0.6}
\begin{table}[h!]
\begin{center}
     \caption{\textcolor{blue}{Notation}}\vspace{-1cm}
    \label{TABLE: table1}
    \begin{tabular}{|M{1cm}|M{6.5cm}|M{1cm}|M{6.5cm}|}\hline
    \small
    Notation & Definition & Notation & Definition\\ \hline
    $\lambda_M$ & Intensity of \ac{TBS} & $\lambda_{UA}$ & Intensity of \ac{UAV-AP} \\\hline
      $\lambda_{UB}$ & Intensity of backhaul connected UAV-BS & $B$ & Bandwidth of the system  \\\hline
      $\theta$ & Polar angle & $\beta$ & Bandwidth Allocation Factor \\\hline
     $g_M$  & Rayleigh fading for TBS in access link & $G_L$ & Nakagami fading for \ac{LoS} UAV-AP in access link\\\hline
     $g_{NL}$  & Rayleigh fading for \ac{NLoS} UAV-AP in access link & $G_B$ & Nakagami fading between UAV-AP and UAV-BS in xHaul link\\\hline
     $g_{B}$  & Rayleigh fading between UAV-AP and TBS in xHaul link & $W_L$ & Probability of \ac{LoS} transmission\\\hline
     $W_N$  & Probability of \ac{NLoS} transmission & $d_{MA}$ & Distance between user-TBS in access link\\\hline
     $d_{UAL}$  &  Distance between user and LoS UAV-AP in access link & $d_{UAN}$ & Distance between the user and NLoS UAV-AP in access link\\\hline
     $d_{MB}$  &  Distance between UAV-AP and TBS in xHaul link & $d_{UB}$ & Distance between the tagged UAV-AP and UAV-BS in xHaul link\\\hline
     $\mathcal{J}$ & Content Database & $L$  & Database size \\\hline
      $\gamma$ & Popularity Index & $C$  & Cache size \\\hline
      $a$ & Content Request Probability & $b$  & Caching probability \\\hline
      $P_{suc}$ & Success Probability & $h$  & Height of tagged \ac{UAV-AP} \\\hline
      $T$ & SINR Threshold & $\Gamma$  & Received SINR at user end \\\hline
      $N_u$ & No.of users in the access link & $R$  & Rate Throughput \\\hline
      $t_a$ & Access SINR threshold & $t_b$  & xHaul SINR threshold \\\hline
     \end{tabular}
     \end{center}
  \end{table}
\vspace{-0.5cm}
\subsection{Channel Model- xHaul link}
We consider that the \acp{UAV-AP} strategically position themselves in the 3D space so as to have an \ac{LoS} visibility state for the xHaul links. For a \ac{UAV-AP} to \ac{UAV-BS} xHaul link, we assume Nakagami distributed fast-fading, $G_{B}$, with parameter $m$~\cite{6}. On the contrary, the fast-fading for the xHaul link between a \ac{TBS} and an \ac{UAV-AP} suffers from a Rayleigh distributed fast-fading, $g_B$ with variance equal to 1, \textcolor{blue}{since Rayleigh fading and Nakagami-m fading offer the same network performance in the strongest \ac{BS} association scheme for a \ac{LoS} transmission.~\cite{34}.} Similar to the access link, we assume a power-law model for the large-scale path loss. Accordingly, the received power at a \ac{UAV-AP} from a \ac{TBS} located at a distance $d_{MB}$ from it is $R_{MB}=K_MP_{M}g_B(d_{MB})^{-\alpha_L}$, where, $\alpha_L$ is the path loss exponent. The received power at a \ac{UAV-AP} from a backhaul connected \ac{UAV-BS} is $R_{UB}= K_UP_{UB}G_B(d_{UB})^{-\alpha_L}$, where, $\alpha_L$ is the path loss exponent.
\vspace{-0.5cm}
\subsection{Caching Strategy}
\textcolor{blue}{The \acp{UAV-AP} are equipped with a local storage capability so as to provide rapid access of popular files to the users. The typical user randomly requests contents from the finite content database stored in \ac{UAV-AP}, $\mathcal{J}= \{f_1,f_2,f_3,.....,f_L\}$, where the database size is $L$.} We assume that each file has the same size, which is normalized to one. A subset of the database is locally cached at the \acp{UAV-AP}. The popularity of the files is modeled according to the Zipf law~\cite{4}. In particular, the popularity or the content request probability of the $i^{th}$ file is given as: $a_i=\frac{i^{-\gamma}}{\sum_{j=1}^L j^{-\gamma}},$
where $\gamma \geq 0$ is the popularity factor.
If the value of $\gamma$ increases, i,e, $\gamma >0$, the trend of popularity of files follows as, $a_i > a_j$ $\forall$ $i > j$.'
For example, if $\gamma$=0, all the files are of equal popularity and if $\gamma > 0$, the files have a decreasing popularity i.e. $a_1 > a_2> a_3...a_L$ where $\sum_{i=1}^L a_i$=1. We assume that the \acp{UAV-AP} can store up to $C$ contents where $C \leq L$~\cite{14}.  
We adopt a probabilistic caching strategy to store the files in the \acp{UAV-AP}. The probability that the $i^{th}$ file is stored in the cache or its caching probability is denoted as $b_i$. Naturally, the caching probability satisfies the condition: $ \sum_{i=1}^L b_i \leq C, 0 \leq b_i \leq 1, \forall i.$
In our scheme, we split the cache size into two parts: the first part of the cache stores the \ac{MPC} and the second part stores the \ac{LPC}. In particular, let us consider that there are $C_0$ \ac{MPC} files. Accordingly, we cache all the \ac{MPC} files, i.e., we set $b_i = 1$, $\forall 1 \leq i \leq C_0$. The remaining $C - C_0$ space of the cache is used to probabilistically cache the remaining $L - C_0$ files by setting:
\vspace{-0.5cm}
\begin{align}
    b_i = \min\left(\frac{a_i\left(C - C_0\right)}{1 - \sum_{j = 1}^{C_0} a_j}, 1\right) \quad \forall C_0 + 1 \leq i \leq C \nonumber 
\end{align}
Then, the cache hit probability at the \ac{UAV-AP}, $\mathcal{P}_{hit}$ is defined as the probability that the requested content by a user is cached in the nearest \ac{UAV-AP}~\cite{5}. This is discussed further in Section~\ref{sec:rate}. A successful content delivery at the user can occur in either of the following possibilities:

\begin{enumerate}
    \item The user is associated to the TBS tier, and the user is under coverage from the nearest TBS. This event is denoted by $S_{t}$.
    \item The user is associated to the \ac{UAV-AP} tier, and:
    \begin{enumerate}
        \item The requested file is cached at the nearest \ac{UAV-AP}, and the user is under coverage from the nearest \ac{UAV-AP}. In this case, the \ac{UAV-AP} delivers the file without xHaul support. This event is denoted as $S_{a}$.
        \item The requested file is not cached at the nearest \ac{UAV-AP}, however, the user is under coverage from it, and the \ac{UAV-AP} is under coverage from either the nearest \ac{TBS} or the nearest backhaul connected \ac{UAV} BS via the xHaul link. This event is denoted as $S_{x}$.
    \end{enumerate}
\end{enumerate}
The success probability $P_{suc}$ is given as
\begin{equation}
    P_{suc}=\mathbb{P}\left(S_{t}\right)+ \mathbb{P}\left(S_{a}\right)+ \mathbb{P}\left(S_{x}\right)
    \label{cacheeq}
\end{equation}
\vspace{-0.25cm}
In this regard, the definition and the characterization of {\it coverage} is discussed in Section IV and the individual terms of the equation above are derived in Section V.

Table~\ref{TABLE: table1} provides the notations used in this paper.


\section{Relevant Distance Distributions and Association Probabilities}
\label{sec:dist}
In this section, we derive the distance distributions of the potential access and the xHaul links. In what follows, based on the tier and the links, we use the subscript triplet $ijk$, where $j \in \{A, B\}$ refers to either the access or backhaul and $i \in \{M, U\}$ refers to either the \ac{TBS} or the \ac{UAV} tier (\ac{UAV-AP} in case $j = A$ and \ac{UAV-BS} in case $j = B$). Furthermore, when $i = U$ and $j = A$, we have $k \in \{L, N\}$ representing the visibility state, i.e., LoS or NLoS. For all other $i$ and $j$, we drop the subscript $k$ for ease of notation.

First, let us note that the distance distribution of the typical user can be derived using the classical result of void probability of a \ac{PPP}~\cite{chiu2013stochastic}:
\vspace{-0.3cm}
\begin{lemma}
The probability density function (pdf) of the distance between the typical user and closest \ac{TBS} in the access link, $d_{MA}$, is given by
\vspace{-0.3cm}
\begin{equation}
    f_{d_{MA}}(x)= 2\pi\lambda_{M} x \exp\left({- \pi \lambda_{M} x^2}\right).
\end{equation}
\end{lemma}
\vspace{-0.3cm}
On the contrary, the \acp{UAV-AP} can be categorized into having either an \ac{LoS} or \ac{NLoS} visibility state. The corresponding distance distributions are presented in the following lemma.
\vspace{-0.3cm}
\begin{lemma}
The pdf of the distances between the typical user and closest \ac{LoS} and the closest \ac{NLoS} \ac{UAV-AP}, denoted by $d_{UAL}$ and $d_{UAN}$, respectively, are given by
\begin{equation}
    f_{d_{UAL}}(x)= 2\pi\lambda_{UA} x^2W'_L\exp{\left(-\frac{2}{3}\pi\lambda_{UA}W'_Lx^3\right)},
\end{equation}
\begin{equation}
    f_{d_{UAN}}(x)=2\pi\lambda_{UA} x^2W'_N\exp{\left(-\frac{2}{3}\pi\lambda_{UA}W'_Nx^3\right)},
\end{equation}
where $W'_L=\int_0^{\frac{\pi}{2}} W_L(\theta)\sin{(\theta)}d\theta$ and $W_L(\theta)$ is the probability of an LoS connection averaged over the polar angle, and $W'_N=\int_0^{\frac{\pi}{2}} \left(1-W_L(\theta)\right)\sin{(\theta)}d\theta$.
\end{lemma}
\vspace{-0.3cm}
\begin{IEEEproof}
See Appendix A.
\end{IEEEproof}

In the xHaul link, the distance distribution between a \ac{UAV-AP} at random height and \ac{UAV-BS}, and between typical \ac{UAV-AP} and \ac{TBS} is presented the following lemma.
\vspace{-0.3cm}
\begin{lemma}
The probability density function of distances between \ac{UAV-AP} and closest \ac{TBS} on the ground, and between \ac{UAV-AP} and \ac{UAV-BS} for xHaul, is denoted by $f_{d_{MB}}$ and $f_{d_{UB}}$ respectively, is given by \big(where $h = d\cos(\theta)$\big):
\begin{align}
    f_{d_{MB}}(x| d, \theta)&=2x\pi\lambda_{M}\exp{\left(-\pi\lambda_{M} (x^2-h^2)\right)}, \; x \geq h \\
    f_{d_{UB}}(x| d, \theta)&=
    \begin{cases}
    f'_{d_{UB}}(x), \quad x \leq h.\\
    f''_{d_{UB}}(x| d, \theta), \quad  x \textgreater h,
    \end{cases}
    \label{cdf:backhaul}
    \end{align}
    \vspace{-0.3cm}
where,
\begin{align}
    f'_{d_{UB}}(x) &= 4\pi\lambda_{UB} x^2 \exp{\left(-\lambda_{UB}\frac{4}{3}\pi x^3\right)}. \nonumber \\
    \vspace{-0.2cm}
    f''_{d_{UB}}(x|d, \theta) &=2\pi \lambda_{UB}\left(x^2+xh\right) \exp{\left[-\lambda_{UB}\left(\frac{2}{3}\pi x^3 + x^2\pi h - \frac{1}{3}\pi h^3 \right)\right]}, \nonumber
\end{align}
where h is the height of the tagged \ac{UAV-AP} from the ground. In case of \ac{LoS} \ac{UAV-AP} association, we have $h = d_{UAL} \cos(\theta)$ and for \ac{NLoS} \ac{UAV-AP} association, $h = d_{UAN} \cos(\theta)$. Here $\theta \sim \mathcal{U}\left[-\frac{\pi}{2}, \frac{\pi}{2}\right]$ is the uniformly distributed random orientation of the tagged \ac{UAV-AP} from the typical user~\textcolor{blue}{\cite{38}}. We use the variable $d_a$ to jointly refer to either $d_{UAL}$ or $d_{UAN}$.
\end{lemma}
\vspace{-0.5cm}
\begin{IEEEproof}
See Appendix B.
\end{IEEEproof}
\vspace{-0.7cm}
\subsection{Access Link Association Probabilities}
As discussed earlier, that in the access link, the typical user can either associate to a \acp{TBS} or an LoS/NLoS \acp{UAV-AP}, based on the maximum power received at the user. For a typical user, the probability of getting associated to a \ac{TBS} is presented in the following lemma.
\vspace{-0.5cm}
\begin{lemma}
The probability that the typical user is associated with the \ac{TBS} in access link, $A_{MA}$, is given by:
\vspace{-0.7cm}
\begin{align}
    A_{MA}=A'_{MA}+A''_{MA},
    \label{asstbs1}
\end{align}
\vspace{-1cm}
\begin{multline}
  \textcolor{blue}{A^{'}_{MA}= \int_0^\infty \int_{C_{M1}} ^\infty \left[\exp{\left(-\frac{2}{3}\pi\lambda_{UA} W'_L {\bigg(\Big(\frac{P_{UA}}{P_{M}}\Big)^{\frac{3}{\alpha_{L}}}w^{\frac{3\alpha_N}{\alpha_{L}}}\bigg)}\right)}- \exp{\left(-\frac{2}{3}\pi\lambda_{UA}W'_Lx^{\frac{3\alpha_{N}}{\alpha_{L}}}\right)}\right]}\\\textcolor{blue}{f_{d_{UAN}}(x)  dx  \exp{\left(-\frac{2}{3}\pi\lambda_{UA}W'_N\bigg({\Big(\frac{P_{UA}}{P_{M}}\Big)^{\frac{3}{\alpha_{N}}}w^{\frac{3\alpha_N}{\alpha_{N}}}\bigg)}\right)}f_{d_{MA}}(w)dw.}
\end{multline}
\vspace{-0.8cm}
\begin{multline*}
    \textcolor{blue}{A^{''}_{MA}=\int_0^\infty \int_{C_{M2}}^\infty \bigg[\exp{\left(-\frac{2}{3}\pi\lambda_{UA} W'_N {\Big(\frac{P_{UA}}{P_{M}}\Big)^{\frac{3}{\alpha_{N}}}w^{\frac{3\alpha_N}{\alpha_{N}}}}\right)}-\exp{\left(-\frac{2}{3}\pi\lambda_{UA}W'_Nx^{\frac{3\alpha_{L}}{\alpha_{N}}}\right)}\bigg]} \\ \textcolor{blue}{f_{d_{UAL}}(x)d(x) \exp{\left(-\frac{2}{3}\pi\lambda_{UA}W'_L\bigg(\Big(\frac{P_{UA}}{P_{M}}\Big)^{\frac{3}{\alpha_{L}}}w^{\frac{3\alpha_N}{\alpha_{L}}}\bigg)\right)}f_{d_{MA}}(w)dw,}
\end{multline*}
where $C_{M1}= (\frac{P_{UA}}{P_{M}})^{\frac{1}{\alpha_{N}}} w^{\frac{\alpha_N}{\alpha_{N}}}$, $C_{M2}=(\frac{P_{UA}}{P_{M}})^{\frac{1}{\alpha_{L}}}w^{\frac{\alpha_N}{\alpha_{L}}}$,


\end{lemma}
\vspace{-0.3cm}
\begin{IEEEproof}
The proof is presented in Appendix C.
\end{IEEEproof}
\textcolor{blue}{Solving with a special case:}
\begin{inparaenum}
    \item \textcolor{blue}{$P_{UA}=P_M$} 
    \item \textcolor{blue}{$\alpha_N=\alpha_L$}
\end{inparaenum}
\begin{equation}
    \textcolor{blue}{A^{'}_{MA} = \frac{W_L^{'}}{W_L^{'} +W_N^{'}}\int_0^\infty \exp{\Bigg(\frac{-2}{3}\pi \lambda_{UA} \Big( W_L^{'} + 2W_{N}^{'}\Big) w^3\Bigg)} f_{d_{MA}}(w) dw}
\end{equation}
\begin{equation}
    \textcolor{blue}{A^{''}_{MA} = \frac{W_N^{'}}{W_L^{'} +W_N^{'}}\int_0^\infty \exp{\Bigg(\frac{-2}{3}\pi \lambda_{UA} \Big( W_N^{'} + 2W_{L}^{'}\Big) w^3\Bigg)} f_{d_{MA}}(w) dw}
\end{equation}



Note that we have averaged out on the distance distribution of the nearest \ac{TBS} from the typical user. However, for a \ac{UAV-AP} association, the association probabilities need to be derived conditioned on the respective access distances because of its impact on the backhaul association. The LoS and NLoS UAV-AP association in the access link is discussed next.
\vspace{-0.5cm}
\begin{lemma}
\label{lemma5}
For a given $d_{UAL}$, the probability that the typical user is associated with the LoS \ac{UAV-AP} in access link, is:
\vspace{-0.3cm}
\begin{equation}
    A_{UAL}(d_{UAL})=A'_{UAL}(d_{UAL})+A''_{UAL}(d_{UAL}),
    \label{asslos2}
\end{equation}
\vspace{-0.8cm}
\begin{equation*}
    A'_{UAL}(d_{UAL}) = \mathbb{P}\left(R_{UAL} > R_{UAN} > R_{MA}|d_{UAL}\right).\
    \end{equation*}
\begin{multline*}
    \textcolor{blue}{A'_{UAL}(d_{UAL})=\int_{C_{L1}}^\infty \bigg[\exp{\Big(-\frac{2}{3}\pi\lambda_{UA} W'_N d_{UAL}^{\frac{3\alpha_{L}}{\alpha_{N}}}\Big)}- \exp\Big(-\frac{2}{3}\pi\lambda_{UA}W'_N\Big(\frac{P_{UA}}{P_{M}}\Big)^{\frac{3}{\alpha_{N}}}x^{\frac{3\alpha_N}{\alpha_{N}}}\Big)\bigg]}\\ \textcolor{blue}{\exp{\Big(-\pi\lambda_{M} {\left(\frac{P_{M}}{P_{UA}}\right)^{\frac{2}{\alpha_N}}d_{UAL}^{\frac{2\alpha_{L}}{\alpha_N}}}\Big)}f_{d_{MA}}(x) dx, }
\end{multline*}
\vspace{-0.5cm}
where $C_{L1}= {(\frac{P_{M}}{P_{UA}})^{\frac{1}{\alpha_N}}d_{UAL}^{\frac{\alpha_{L}}{\alpha_N}}}$.
\begin{equation*}
    A''_{UAL}(d_{UAL}) = \mathbb{P}\left(R_{UAL} > R_{MA} > R_{UAN}|d_{UAL}\right).
\end{equation*}
\begin{multline*}
  \textcolor{blue}{A''_{UAL}(d_{UAL}) =\int_{d_{UAL}^{\frac{\alpha_{L}}{\alpha_{N}}}}^\infty \bigg[\exp\left(-\pi\lambda_{M}\left(\frac{P_{M}}{P_{UA}}\right)^{\frac{2}{\alpha_N}}d_{UAL}^{\frac{2\alpha_{L}}{\alpha_N}}\right)-
    \exp\left(-\pi\lambda_{M}\left(\frac{P_{M}}{P_{UA}}\right)^{\frac{2}{\alpha_N}}x^{\frac{2\alpha_{N}}{\alpha_N}}\right)\bigg]}\\ \textcolor{blue}{\exp\Bigg(-\frac{2}{3}\pi\lambda_{UA}W'_Nd_{UAL}^{\frac{3\alpha_{L}}{\alpha_{N}}}\Bigg)f_{d_{UAN}}(x)dx.}
\end{multline*}
   
\end{lemma}
\begin{IEEEproof}
The proof is presented in Appendix D.
\end{IEEEproof}
\vspace{-0.2cm}
We note that the probability of \ac{LoS} \ac{UAV-AP} association in the access link, de-conditioned on $d_{UAL}$ is given by:
\vspace{-0.8cm}
\begin{align}
    \bar{A}_{UAL} =    \int_0^{\infty} A_{UAL}(x) f_{d_{UAL}}(x) dx.
    \label{losass2}
    \nonumber
\end{align}

The above expression prescribes required deployment densities of the \acp{UAV-AP}. Interestingly, in order to maximize the LoS UAV association, extreme densification can be detrimental as discussed below:
\vspace{-0.5cm}
\begin{proposition}
$\bar{A}_{UAL} \to 0$ as $\lambda_{UA} \to \infty$ and has at least one maxima with respect to $\lambda_{UA}$. Accordingly, there exists optimal UAV densities which maximizes the probability of association of typical user with the \ac{LoS} \ac{UAV-AP}.
\end{proposition}
\vspace{-0.5cm}
\begin{proof}
See Appendix E.
\end{proof}
\vspace{-0.5cm}
\begin{lemma}
For a given $d_{UAN}$, the probability that the typical user is associated with the NLoS \ac{UAV-AP} in access link, is:
\vspace{-0.8cm}
\begin{align}
    A_{UAN}=A'_{UAN}(d_{UAN}) +A''_{UAN}(d_{UAN}).
\end{align}
\begin{equation*}
    A'_{UAN}(d_{UAN})= \mathbb{P}\left(R_{UAN} > R_{M} > R_{UAL}\right).
\end{equation*}
\begin{multline}
    = \int_{C_{N1}}^\infty \bigg[\exp{\Big(-\frac{2}{3}\pi\lambda_{UA} W'_L d_{UAN}^{\frac{3\alpha_{N}}{\alpha_{L}}}}\Big)-\exp{\Big(-\frac{2}{3}\pi\lambda_{UA}W'_L (\frac{P_{UA}}{P_{M}})^{\frac{3}{\alpha_{L}}}x^{\frac{3\alpha_N}{\alpha_{L}}}}\Big)\bigg]\\ \exp{\left(-\pi\lambda_{M} (C_{N1})^2\right)}f_{d_{MA}}(x)dx. \nonumber
\end{multline}
\vspace{-1cm}
\begin{equation*}
    A''_{UAN}(d_{UAN})= \mathbb{P}\left(R_{UAN} > R_{UAL} > R_M\right). 
\end{equation*}
\begin{multline}
    =\int_{d_{UAN}^{\frac{\alpha_{N}}{\alpha_{L}}}}^\infty \bigg[\exp{\Big(-\pi\lambda_{M}(N_2)^2\Big)}-\exp{\Big(-\pi\lambda_{M}(N_3)^2\Big)}\bigg]\\
    \exp{\Big(-\frac{2}{3}\pi\lambda_{UA}W'_L d_{UAN}^{\frac{3\alpha_{N}}{\alpha_{L}}}\Big)}f_{d_{UAL}}(x)dx,
\end{multline}
where $C_{N1}=(\frac{P_{M}}{P_{UA}})^{\frac{1}{\alpha_N}}d_{UAN}^{\frac{\alpha_{N}}{\alpha_N}}$, $N_2=(\frac{P_{M}}{P_{UA}})^{\frac{1}{\alpha_N}}d_{UAN}^{\frac{\alpha_{N}}{\alpha_N}}$, $N_3=(\frac{P_{M}}{P_{UA}})^{\frac{1}{\alpha_N}}x^{\frac{\alpha_{L}}{\alpha_N}}$.
\end{lemma}
\vspace{-0.3cm}
As a result, the probability of \ac{NLoS} \ac{UAV-AP} association in the access link, de-conditioned on $d_{UAL}$ is given by:
\vspace{-0.8cm}
\begin{align}
    \bar{A}_{UAN} =  \int_0^{\infty} A_{UAN}(x) f_{d_{UAN}}(x) dx. \nonumber
\end{align}
It can be verified with algebraic manipulations as well as numerically (as discussed in Section VI), that $A_{MA} + \bar{A}_{UAL} + \bar{A}_{UAN} = 1$.
\vspace{-0.2cm}
\begin{corollary}
When taking into account the height of \acp{UAV-AP}, increasing the height of \ac{LoS} \ac{UAV-AP} will significantly decrease the association of typical user with the \ac{LoS} \ac{UAV-AP}.
\end{corollary}
\vspace{-0.5cm}
\begin{proof}
\textcolor{blue}{According to the RSSI based association scheme, the typical user connects to the \ac{LoS} \ac{UAV-AP} tier when either of the following events are true:}

\textcolor{blue}{(i) $R_{UAL} > R_{UAN} > R_{MA}$} \textcolor{blue}{(ii) $R_{UAL} > R_{MA}> R_{UAN}$}

\textcolor{blue}{The probability of event (i) can be written as}
\begin{equation}
     \textcolor{blue}{\mathbb{P}\Big(P_{UA}d_{UAL}^{-\alpha_{L}} > P_{UA}d_{UAN}^{-\alpha_{N}} > P_{M}d_{MA}^{-\alpha_{N}}\Big).}
     \label{asslos}
\end{equation}
\textcolor{blue}{Representing the distance between user and \ac{LoS} UAV-AP, $d_{UAL}$ in terms of height of \ac{LoS} UAV-AP, $h_L$ and $\theta$,}
\vspace{-0.6cm}
\begin{equation}
     \textcolor{blue}{\mathbb{P}\Big(P_{UA}\Big(\frac{h_L}{\cos{(\theta)}}\Big)^{-\alpha_{L}} > P_{UA}d_{UAN}^{-\alpha_{N}} > P_{M}d_{MA}^{-\alpha_{N}}\Big).}
     \label{asslos}
\end{equation}
\textcolor{blue}{Here $\theta \sim \mathcal{U}\left[-\frac{\pi}{2}, \frac{\pi}{2}\right]$ is the uniformly distributed random orientation of the tagged \ac{UAV-AP} from the typical user.
Taking expectation over $d_{MA}$ conditioning on $\theta$, the final expression written as} 
\vspace{-0.4cm}
\begin{multline}
  \textcolor{blue}{\mathbb{P}\left(R_{UAL} > R_{UAN} > R_{MA}|\theta\right)= \int_{C_{L1}}^\infty \bigg[\exp{\Bigg(-2\pi\lambda_{UA} W'_N \int_0^{\Big(\frac{h_L}{\cos{(\theta)}}\Big)^{\frac{\alpha_{L}}{\alpha_{N}}}}z^2dz\Bigg)}-} \\ \textcolor{blue}{\exp\Big(-2\pi\lambda_{UA}W'_N\int_0^{L_1}y^2 dy\Big)\bigg] \exp{\Big(-\pi\lambda_{M} (C_{L1})^2\Big)}f_{d_{MA}}(x) dx, }
  \label{eq1}
\end{multline}
\textcolor{blue}{where $C_{L1}= {(\frac{P_{M}}{P_{UA}})^{\frac{1}{\alpha_N}}\Big(\frac{h_L}{\cos{(\theta)}}\Big)^{\frac{\alpha_{L}}{\alpha_N}}},  L_1=(\frac{P_{UA}}{P_{M}})^{\frac{1}{\alpha_{N}}}x^{\frac{\alpha_N}{\alpha_{N}}}$
and }
\begin{multline}
  \textcolor{blue}{\mathbb{P}\left(R_{UAL} > R_{MA} > R_{UAN}|\theta\right) =\int_{\Big(\frac{h_L}{\cos{(\theta)}}\Big)^{\frac{\alpha_{L}}{\alpha_{N}}}}^\infty \bigg[\exp\left(-\pi\lambda_{M}\left(L_2\right)^2\right)-
    \exp\left(-\pi\lambda_{M}\left(L_3\right)^2\right)\bigg] } \\ \textcolor{blue}{\exp\Bigg(-2\pi\lambda_{UA}W'_N\int_0^{\Big(\frac{h_L}{\cos{(\theta)}}\Big)^{\frac{\alpha_{L}}{\alpha_{N}}}}x^2 dx\Bigg)f_{d_{UAN}}(x)dx ,}
    \label{eq2}
\end{multline}
\textcolor{blue}{where $L_2=(\frac{P_{M}}{P_{UA}})^{\frac{1}{\alpha_N}}\Big(\frac{h_L}{\cos{(\theta)}}\Big)^{\frac{\alpha_{L}}{\alpha_N}}, L_3=(\frac{P_{M}}{P_{UA}})^{\frac{1}{\alpha_N}}x^{\frac{\alpha_{N}}{\alpha_N}}$.}

\textcolor{blue}{Adding (\ref{eq1}) and (\ref{eq2}), we get the probability of \ac{LoS} \ac{UAV-AP} association in the access link $A_{UAL}$, de-conditioned on $\theta$ is given by: $\textcolor{blue}{\bar{A}_{UAL} (h_L) =    \int_{-\pi/2}^{\pi/2} A_{UAL}(\rho) f_{\theta}(\rho) d(\rho)}$}.
\textcolor{blue}{where $f_{\theta}(\rho)= \frac{1}{\pi}$.
So, as we increase the height of \ac{LoS} \ac{UAV-AP} $h_L$, the \ac{LoS} association probability $\bar{A}_{UAL}$ decreases. This is because, as $h_L$ increases, the received power from the \ac{LoS} \ac{UAV-AP} at the typical user decreases. Thus the probability of associating to \ac{NLoS} \ac{UAV-AP}/\ac{TBS} increases.}

\textcolor{blue}{Further, we can derive the probability of associating to \ac{NLoS} UAV-AP, $\bar{A}_{UAN}$, by keeping the distance of \ac{LoS} UAV-AP from the typical user in terms of $h_L$ and $\theta$. Furthermore, we can derive the probability of associating to \ac{TBS} $\bar{A}_M$.}
\end{proof}
\vspace{-1cm}
\subsection{xHaul Link Association Probabilities}
\vspace{-0.2cm}
Next, we consider the xHaul link in case of a \acp{UAV-AP} association in the access link. As discussed before, the xHaul association probabilities are dependent on the access link distance, $d_a$ of the tagged \ac{UAV-AP} from the typical user. 
Depending on the visibility state of the tagged \ac{UAV-AP} from the typical user, the $d_a$ can either be $d_{UAL}$ or $d_{UAN}$. The tagged \ac{UAV-AP} associates with either the \ac{TBS} tier or a backhaul connected \ac{UAV-BS} tier for xHaul support. Similar to the access link, the xHaul association is also based on RSSI measurements. The probabilities are given in the following lemma.
\vspace{-0.5cm}
\begin{lemma}
The probability, $A_{UB}(d_a, \theta)$ that the tagged \ac{UAV-AP} associates to the \ac{UAV-BS} tier for xHaul support is given as:
\begin{multline}
    A_{UB}(d_a, \theta) = \int_{d_a \cos(\theta)}^{\ell(d_a, \theta)} 1 - \exp\left(-\frac{4}{3}\pi \lambda_{UB} \left(\frac{P_M}{P_{UB}} x^{-\alpha_L}\right)^{-\frac{3}{\alpha_L}}\right) f_{d_{MB}}(x)dx  + \\ \int_{\ell(d_a, \theta)}^\infty F''_{d_{UB}}\left(\left(\frac{P_M}{P_{UB}} d_{MB}^{-\alpha_L}\right)^{-\frac{1}{\alpha_L}}\right) f_{d_{MB}}(x)dx,
\end{multline}
where, 
\begin{gather}
    \ell(d_a, \theta) =  \left(\left(d_a \cos(\theta)\right)^{-\alpha_L}\frac{P_{UB}}{P_M}\right)^{-\frac{1}{\alpha_L}}, \nonumber
\end{gather}
\begin{equation}
    F''_{d_{UB}}(x)
    =1- \exp{\bigg[-\lambda_{UB}\left(\frac{4}{3}\pi x^3 - \frac{1}{3}\pi\big[2x^3-3x^2h+h^3\big] \right)\bigg]}
    \nonumber 
\end{equation}
is the CDF corresponding to the pdf $f''_{d_{UB}}(\cdot)$ given in \eqref{cdf:backhaul}.
Naturally, the probability, $A_{MB}(d_a, \theta)$ that the tagged \ac{UAV-AP} associates to the \ac{TBS} tier for xHaul support is given as $A_{MB}(d_a, \theta) = 1 - A_{UB}(d_a, \theta)$. 
\end{lemma}
\vspace{-0.3cm}
\begin{IEEEproof}
See Appendix F.
\end{IEEEproof}
\vspace{-0.5cm}
\section{Characterization of SINR Coverage Probability}
\label{sec:SINR}
The SINR coverage probability for the typical user, $\mathcal{P}_C$ is defined as the probability that the received \ac{SINR} $\Gamma$ at the typical user is greater than a \ac{SINR} threshold $T$, i.e., $\mathcal{P}_C= \mathbb{P}( \Gamma \textgreater T)$.
\vspace{-0.1cm}
Ergodically, this represents the fraction of users that are under SINR coverage from the network. In the following, we adapt the same notation as above and denote the instantaneous SINR and the corresponding coverage probability by $\Gamma_{ijk}$ and $\mathcal{P}_{ijk}$, respectively.
\vspace{-0.5cm}
\begin{lemma}
The \ac{SINR} coverage probability of the typical user associated to the \ac{TBS} in the access link is given as
\begin{equation}
   \mathcal{P}_{CM}(t_a) =\int_0^\infty \left( \exp{\left(\frac{-t_a N_0}{K_MP_{M}q^{-\alpha_N}}\right)}  I'_{M1}  I'_{L1}  I'_{NL1} \right) f_{d_{MA}}(q)dq,
   \label{sinrmbs}
\end{equation}
where $I'_{M1}$, $I'_{L1}$ and $I'_{NL1}$ are the interference terms from other \acp{TBS}, \ac{LoS} and \ac{NLoS}-\acp{UAV}, respectively. $t_a$ is the \ac{SINR} threshold in the access link and $f_{d_{MA}}(q)$ is the pdf of the distances between user and closest TBS.
\begin{equation}
    I'_{M1}=\exp{\left(-2\pi\lambda_{M} \int_{d_{MA}}^\infty \left(1-\frac{1}{1+\frac{t_{a} t^{-\alpha_N}}{q^{-\alpha_N}}}\right)t dt\right)}.
    \label{intertbs}
\end{equation}

\begin{equation}
    I'_{L1}= \exp\left(-2\pi\lambda_{UA} W'_L\int_{J_{L}}^{\infty} \Bigg[x^{2} dx - x^{2} \left(\frac{m}{m+\frac{\eta t_{a} K_{U}P_{UA}x^{-\alpha_{L}}}{{{K_{M}P_{M}q^{-\alpha_N}}}}} \right)^{m} dx \Bigg]\right),
    \label{sinrlosinterference}
\end{equation}

where $J_{L}=(\frac{P_{UA}}{P_{M}})^{\frac{1}{\alpha_{L}}} q^{\frac{\alpha_N}{\alpha_{L}}}, W'_L= \int_0^{\frac{\pi}{2}} W_L(\theta)  \sin{(\theta)} d\theta$ and $\eta= m(m!)^{\frac{-1}{m}}$.

\begin{equation}
    I'_{NL1}= \exp\left(-2\pi\lambda_{UA} W'_N\int_{J_{NL}}^\infty \Bigg [ y^2 dy - y^2 \left(\frac{1}{1+\frac{t_{a} K_UP_{UA}y^{-\alpha_{N}}}{{K_MP_{M}q^{-\alpha_N}}}}\right) dy \Bigg] \right), \nonumber
\end{equation}

where $J_{NL}={(\frac{P_{UA}}{P_{M}})^{\frac{1}{\alpha_{N}}} q^{\frac{\alpha_N}{\alpha_{N}}}}$, $W'_N=\int_0^{\frac{\pi}{2}} \left(1-W_L(\theta)\right)\sin{(\theta)}d\theta$.
\end{lemma}
\vspace{-0.5cm}
\begin{lemma}
Given an \ac{NLoS} and \ac{LoS} \ac{UAV-AP} association, the \ac{SINR} coverage probability of the typical user in the access link is given respectively as:
\subsubsection{NLoS \ac{UAV-AP}}
\begin{equation}
   \mathcal{P}_{CN}(d_{UAN}, t_a) = \left( \exp{\left(\frac{-t_a N_0}{K_UP_{UA}d_{UAN}^{-\alpha_{N}}}\right)} \cdot I'_{M2} \cdot I'_{L2} \cdot I'_{NL2} \right),
   \label{sinrnlos}
\end{equation}
$I'_{NL2}$, $I'_{L2}$ and $I'_{M2}$ are interference terms from other \ac{NLoS}, \ac{LoS} \acp{UAV-AP} and \acp{TBS} respectively.
\begin{equation}
    I'_{M2}=\exp\Big(-2\pi\lambda_{M} \int_{J_M'}^\infty \Big(1- \frac{1}{1+\frac{t_{a} K_MP_{M} t^{-\alpha_N}}{K_UP_{UA}d_{UAN}^{-\alpha_{N}}}}\Big)t dt\Big), \nonumber
\end{equation}
\begin{equation}
   I'_{L2}=\exp\left(-2\pi\lambda_{UA} W'_L\int_{ J_L'}^\infty \Big[ y^2 dy -  y^2 \left(\frac{m}{m+\frac{\eta t_{a} y^{-\alpha_{L}}}{{{d_{UAN}^{-\alpha_{N}}}}}}\right)^m dy \Big]\right), \nonumber
\end{equation}\vspace{-0.2cm}
\begin{equation}
   I'_{NL2}=\exp\bigg(-2\pi\lambda_{UA} W'_N\int_{d_{UAN}}^\infty\Big(1- \frac{1}{1+\frac{t_a x^{-\alpha_{N}}}{{d_{UAN}^{-\alpha_{N}}}}}\Big)x^2dx\bigg), \nonumber
\end{equation}
where $J_M'={\frac{P_{M}}{P_{UA}}^{\frac{1}{\alpha_N}}d_{UAN}^{\frac{\alpha_{N}}{\alpha_N}}}$ and $J_L'=d_{UAN}^{\frac{\alpha_{N}}{\alpha_{L}}}.$ 
\subsubsection{LoS \ac{UAV-AP}}
\begin{equation}
    \mathcal{P}_{CL}(d_{UAL}, t_a) =\sum_{n=1}^m (-1)^{n+1} \Comb{m}{n} \exp{\left(\frac{-n \eta t_a N_0}{K_UP_{UA}d_{UAL}^{-\alpha_{L}}}\right)} I'_{M3}  I'_{L3}  I'_{NL3},  
    \label{sinrlos}
\end{equation}
$I'_{L3}$, $I'_{NL3}$ and $I'_{M3}$ are interference terms from other \ac{LoS}, \ac{NLoS} \acp{UAV-AP} and \acp{TBS} respectively. 
\begin{equation}
    I'_{M3}= \exp\left(-2\pi\lambda_{M} \int_{J_M^{''}}^\infty \Big[ t dt - t \left( \frac{1}{1+\frac{n \eta t_a K_MP_{M} t^{-\alpha_N}}{K_UP_{UA}d_{UAN}^{-\alpha_{L}}}}\right) dt \right), \nonumber
\end{equation}
\begin{equation}
    I'_{L3}= \exp\left(-2\pi\lambda_{UA} W'_L\int_{ d_{UAL}}^\infty \Big[ x^2 dx - x^2\left(\frac{m}{m+\frac{n \eta t_a x^{-\alpha_{L}}}{{{d_{UAL}^{-\alpha_{L}}}}}}\right)^m dx \Big]\right),
    \label{interflos}
\end{equation}
\begin{equation}
    I'_{NL3}= \exp\left(-2\pi\lambda_{UA} W'_N\int_{J_{NL}^{''}}^\infty \Big[ y^2 dy - y^2 \left( \frac{1}{1+\frac{n \eta t_a y^{-\alpha_{N}}}{{d_{UAL}^{-\alpha_{L}}}}}\right) dy \Big] \right), \nonumber
\end{equation}
where $J_M^{''}= \frac{P_{M}}{P_{UA}}^{\frac{1}{\alpha_N}}d_{UAL}^{\frac{\alpha_{L}}{\alpha_N}}$ and $J_{NL}^{''}=d_{UAL}^{\frac{\alpha_{L}}{\alpha_{N}}}.$
\end{lemma}
\begin{IEEEproof}
See Appendix G.
\end{IEEEproof}

\vspace{-0.5cm}
\begin{lemma}
The \ac{SINR} coverage probability of the tagged \ac{UAV-AP} associated to the \ac{TBS} in the xHaul link is given as
\begin{equation}
    \mathcal{P}_{CMB} (d_{MB}, t_b)=  \exp{\left(\frac{-t_b N_0}{K_MP_{M}(d_{MB})^{-\alpha_L}}\right)} \cdot I'_{M4} \cdot I'_{U4}, 
    \label{sinrx1}
\end{equation}
$I'_{M4}$ and $I'_{U4}$ are interference terms from other \acp{TBS} and \acp{UAV-BS} respectively. $t_b$ is the SINR threshold in the xHaul link.
\begin{equation}
    I'_{M4}= \exp{\left(-2\pi\lambda_{M} \int_{d_{MB}}^\infty \left(1-\frac{1}{1+\frac{t_b t^{-\alpha_L}}{(d_{MB})^{-\alpha_L}}}\right)t dt\right)}. \nonumber
\end{equation}
\begin{equation}
    I'_{U4}= \exp\left(-2\pi\lambda_{UB} W'_L\int_{J_{BM}^{'}}^\infty \Big[ x^2 dx - x^2 \left(\frac{m}{m+\frac{t_b K_UP_{UB}x^{-\alpha_L}}{{{K_MP_{M}d_{MB}^{-\alpha_L}}}}}\right)^m dx \Big]\right), \nonumber
\end{equation}
where $J_{BM}^{'}=(\frac{P_{UB}}{P_{M}})^{\frac{1}{\alpha_L}} d_{MB}^{\frac{\alpha_L}{\alpha_L}}.$
\end{lemma}
\vspace{-0.8cm}
\begin{lemma}
The \ac{SINR} coverage probability of the typical \ac{UAV-AP} associated to the \ac{UAV-BS} in the xHaul link is:
\vspace{-0.5cm}
\begin{multline}
    \mathcal{P}_{CUB}(d_{UB}, t_b) = \textcolor{blue}{\sum_{n=1}^m (-1)^{n+1} \cdot \Comb{m}{n} \exp{\left(\frac{-n \eta t_b N_0}{K_UP_{UB}d_{UB}^{-\alpha_L}}\right)}} \cdot I'_{M5}  I'_{H5}   + \\ \sum_{n=1}^m (-1)^{n+1} \cdot \Comb{m}{n} \exp{\left(\frac{-n \eta t_b N_0}{K_UP_{UB}d_{UB}^{-\alpha_L}}\right)}  I'_{M5}  I'_{U5},
    \label{sinrx2}
\end{multline}
$I'_{U5}$ and $I'_{M5}$ are interference terms from other \acp{UAV-BS} and \acp{TBS} respectively.
\vspace{-0.3cm}
\begin{equation}
    I'_{M5}= \exp\left(-2\pi\lambda_{M} \int_{J_{BU}'}^\infty \Big[ t dt - t \left(\frac{1}{1+\frac{n \eta t_b K_MP_{M} t^{-\alpha_L}}{K_UP_{UB} d_{UB}^{-\alpha_L}}}\right)dt\Big]\right), \nonumber
\end{equation}
where $J_{BU}'= \frac{P_{M}}{P_{UB}}^{\frac{1}{\alpha_L}}d_{UB}^{\frac{\alpha_L}{\alpha_L}}.$
\vspace{-0.3cm}
\begin{equation}
    I'_{U5}= \exp\left(-2\pi\lambda_{UB} W'_L\int_{d_{UB}}^\infty \Big [ x^2 dx - x^2 \left(\frac{m}{m+\frac{n \eta t_b x^{-\alpha_L}}{{{d_{UB}^{-\alpha_L}}}}}\right)^m dx \Big]\right).
    \label{interfxHaul}
\end{equation}
\end{lemma}
\section{Rate Coverage and Content Delivery Success}
\label{sec:rate}
The framework for SINR coverage probability can be employed to derive the rate coverage probability, which is defined as the probability that the per-user rate at the typical user is greater than a given threshold $r_0$. Mathematically, for $N_u$ simultaneous users in the access link with orthogonal channel allocation, we have:
\begin{gather}
    \mathbb{P}\left(R \geq r_0\right)= \mathbb{P}\left(\frac{\beta B}{N_u}\log_2\left(1 + \Gamma\right) \geq r_0\right) 
    = \mathbb{P}\left(\Gamma \geq 2^{\frac{N_u r_0}{\beta B}} - 1\right) 
    = \mathcal{P}_C\left(2^{\frac{N_u r_0}{\beta B}} - 1\right) 
    \label{rate_cov}
\end{gather}
Let us assume that the minimum rate requirement in the access link to transfer a file requested by the user before the service deadline be given by $r_a$. Accordingly, for a successful transmission, the access link must sustain an SINR above a threshold given by: 
$t_a= 2^{\frac{N_ur_a}{B\beta}-1}.$ 
Similarly, for an xHaul rate threshold of $r_b$, we have the following xHaul SINR threshold:
    $t_b= 2^{\frac{r_b}{B(1-\beta)}-1}.$ 
Additionally, the cache hit probability or the probability that the requested the file is stored in the cache, $\mathcal{P}_{hit}$ is:
   $ \mathcal{P}_{hit}= \sum_{i=1}^{C} a_i b_i.$ 
Consequently, the probability that the requested file is not stored in the cache, $\mathcal{P}_{miss}$ is:
    $\mathcal{P}_{miss}= \sum_{i=1}^{C} a_i  (1-b_i).$ 
The successful content delivery of the user in (\ref{cacheeq}) is defined as $\mathbb{P}(S_t) = A_{MA} \mathcal{P}_{CM}\left(t_a\right) $ and 
\vspace{-0.5cm}
\begin{multline}
     \mathbb{P}(S_a) = \mathcal{P}_{hit} \left(\textcolor{blue}{\underbrace{\int_0^\infty A_{UAL}(x)\mathcal{P}_{CL}(x, t_a) f_{d_{UAL}}(x) dx}_{\RomanNumeralCaps{1}}+ \underbrace{\int_0^\infty A_{UAN}(x)\mathcal{P}_{CN}(x, t_a) f_{d_{UAN}}(x) dx}_{\RomanNumeralCaps{2}}}\right)
\end{multline}
\vspace{-1cm}
 \begin{multline}
     \mathbb{P}(S_x) =  \mathcal{P}_{miss}\Bigg[ \int_{-\frac{\pi}{2}}^{\frac{\pi}{2}}\Bigg(\int_0^{\infty}\left(A_{UAL}(x)\mathcal{P}_{CL}(x, t_a) \mathit{B_l}\left(x, \theta\right) \right) f_{UAL}(x) dx d\theta \Bigg) +\\ \int_{-\frac{\pi}{2}}^{\frac{\pi}{2}}\Bigg(\int_0^\infty A_{UAN}(x)\mathcal{P}_{CN}(x, t_a)\mathit{B_l}(x, \theta)f_{UAN}(x)dx d\theta \Bigg)\Bigg]
 \end{multline} 
 \vspace{-0.5cm}
where,
\begin{multline*}
    \mathit{B_l}(x, \theta) = \int_0^{\infty} A_{UB}(x, \theta)\mathcal{P}_{CUB}(y, t_b)f_{d_{UB}}(y|x,\theta) dy + \\ 
    \int_{x \cos(\theta)}^\infty A_{MB}(x, \theta)\mathcal{P}_{CMB}(x, t_b) f_{d_{MB}}(y|x, \theta)dy \nonumber
\end{multline*}
where $\mathit{B_l}(x,\theta)$ is the total xHaul coverage probability. Here $\RomanNumeralCaps{1}$ is the \ac{SINR} coverage probability of the user associated to \ac{LoS} \ac{UAV-AP} and $\RomanNumeralCaps{2}$ is the \ac{SINR} coverage probability of the user associated to \ac{NLoS} \ac{UAV-AP}.
Thus, we have characterized the different components of the expression \eqref{cacheeq} which characterize the content delivery success.
\vspace{-0.5cm}
\section{Results and discussions}
\label{sec:result}
In this section, we validate our analytical framework using Monte-Carlo simulations, \textcolor{blue}{ providing a precise analysis equivalent to executing experiments,} and present some numerical results to discuss the salient features of the network. The transmit powers are $P_{UA}$=27dBm~\cite{3}, $P_{UB}$= 33 dBm~\cite{1} and $P_M$=46 dBm~\cite{13}. The thresholds are $r_a$= 1.1 Mbps~\cite{28} and $r_b$= 80 Mbps~\cite{27}. $\alpha_L$=2, $\alpha_N$=4, $L$=1000~\cite{7} and $B$= 100 MHz~\cite{27}.

\vspace{-0.5cm}  
\subsection{Association Probabilities and Validation of SINR Coverage}
   \begin{figure}[htbp]
   \begin{minipage}[t]{0.45\textwidth}
    \centering
    \includegraphics[width=1.05\linewidth,height=0.8\linewidth]{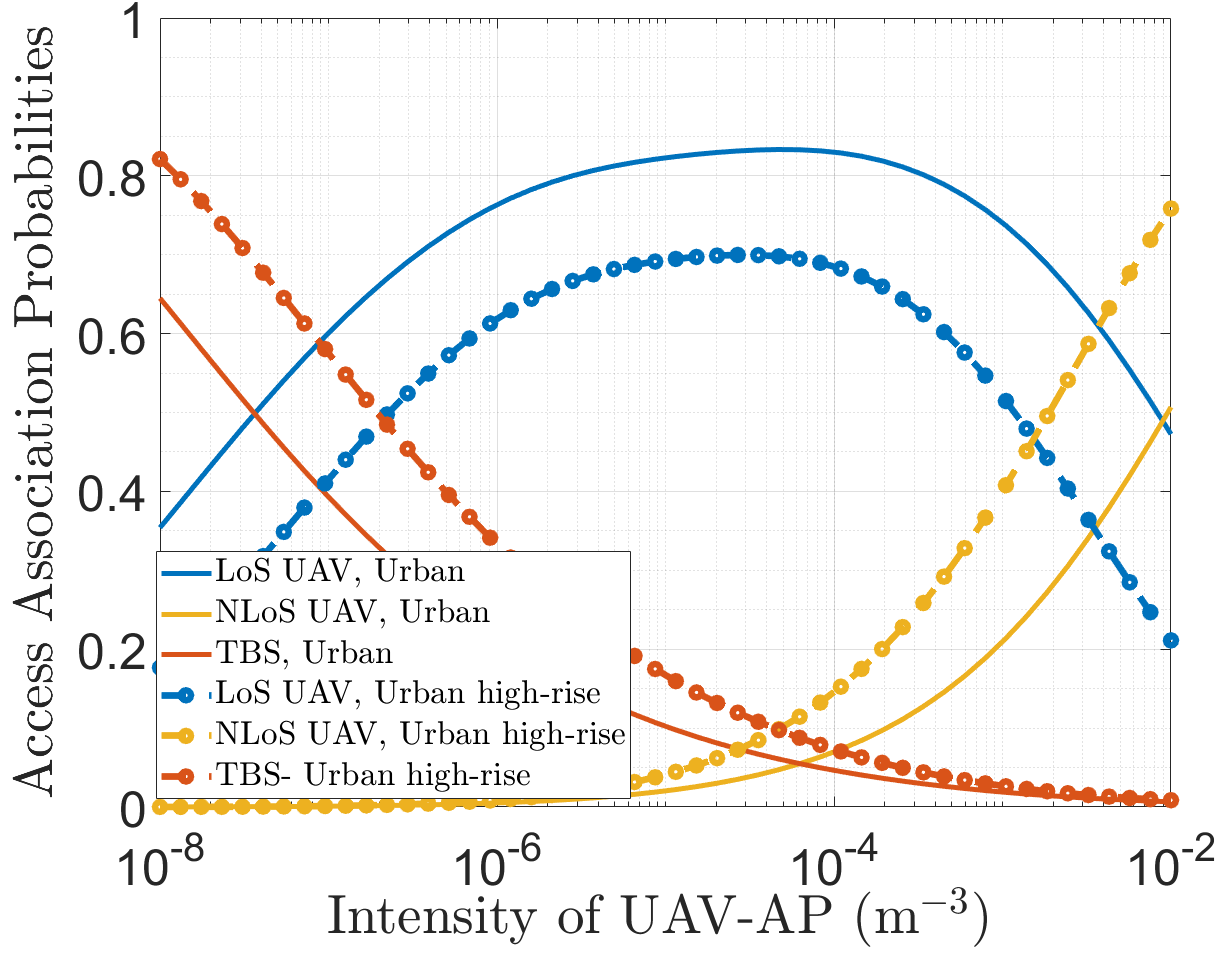}
    \caption{Access association probabilities versus intensity of \acp{UAV-AP} for the typical user in the access link for $\lambda_{M}=10^{-6} m^{-2}$.}
    \label{fig:access_asso}
    \end{minipage}\hfill
    \begin{minipage}[t]{0.45\textwidth}
     \centering
    \includegraphics[width=1.05\linewidth]{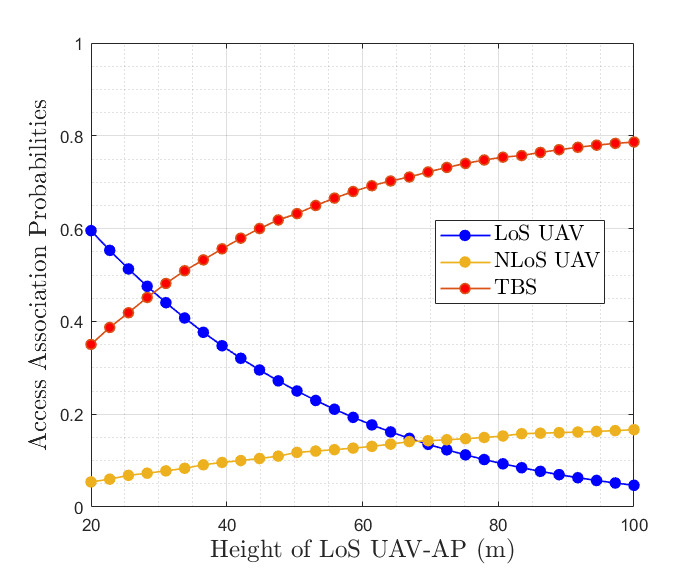}
    \caption{Access association probabilities versus height of LoS UAV-AP}
    \label{fig:acc_ass_height}
    \end{minipage}
    \end{figure}
    
In Fig.~\ref{fig:access_asso}, we plot the association probabilities of a typical user in the access link versus the intensity of \ac{UAV-AP} for urban and high-rise urban scenarios.The environment parameters chosen for urban and high-rise urban are $a$ = 9.61, $b$ = 0.61 and $a$ = 24.23, $b$ = 0.08 respectively.
As discussed in Lemma~{\ref{lemma5}}, we note that the probability of LoS UAV-AP association increases with $\lambda_{UA}$, reaches a maximum, and decreases with further increase in $\lambda_{UA}$. 
This is because a fractional increase in the density increases the number of Los links for the very sparse deployment of UAVs. However, beyond a certain density, increasing the number of UAVs in the network further increases the potential of serving NLoS UAV-APs without substantially increasing the number of LoS links.We note that the intensity that maximizes the LOS UAV-AP association is lower for the urban environment than the urban high-rise environment due to larger blockage sizes. Intuitively, the \ac{TBS} association decreases with an increasing number of UAVs in the network. \textcolor{blue}{Also, due to RSSI based association scheme, the access association probability does not depend on the intensity of users but only on the received powers from different tiers.} Thus, this analysis equips the operator with an essential insight: in order to maximize LoS connectivity densification of the network does not help beyond a certain limit. Accordingly, our analysis prescribes the optimal deployment density to maximize the LoS association for a given blockage environment.
\begin{figure}[!htb]
\begin{minipage}[t]{0.45\textwidth}
     \centering
    \includegraphics[width=1.05\linewidth]{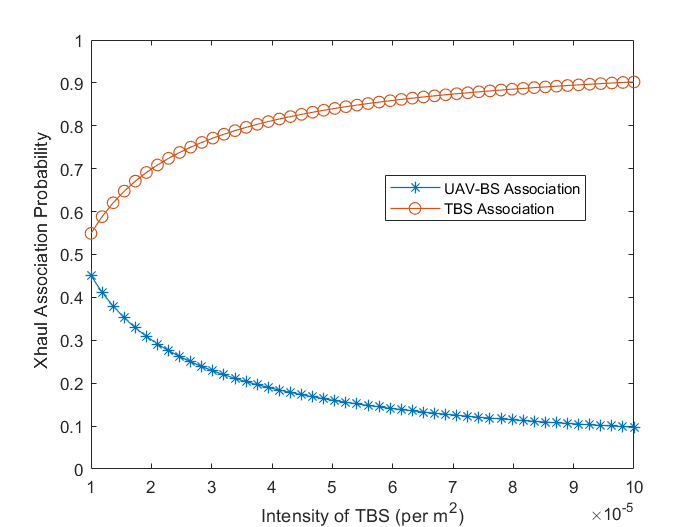}
    \caption{xHaul association probabilities versus intensity of \ac{TBS} for a typical \ac{UAV-AP} in the xHaul link for $\lambda_{UA}=10^{-5} m^{-3}$ and $\lambda_{UB}=10^{-7}m^{-3}$.}\label{fig:x-Haul_asso}
    \end{minipage}\hfill
    \begin{minipage}[t]{0.45\textwidth}
    \centering
    \includegraphics[width=1.05\linewidth,height=0.8\linewidth]{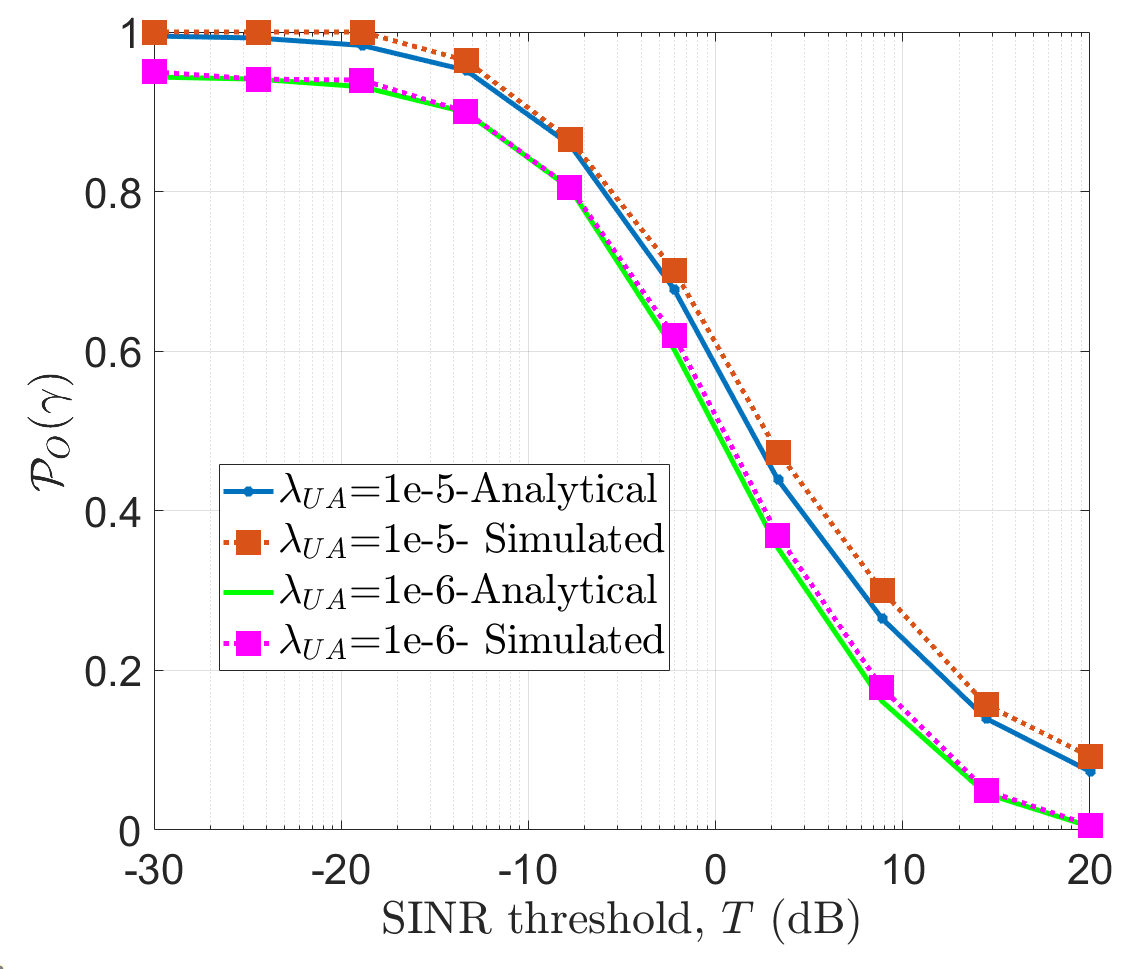}
    \caption{Overall coverage probability of the network versus \ac{SINR} threshold for $\lambda_{M}=10^{-6} m^{-2}$, $\lambda_{UA}=10^{-5} m^{-3}$, $\eta =9.61$, $\mu =0.16$.}\label{fig:overall_cov}
    \end{minipage}
    \end{figure}
\textcolor{blue}{In Fig.~\ref{fig:acc_ass_height}, we plot the association probabilities of a user in the access link versus the height of \ac{LoS} \ac{UAV-AP}. We observe that as the height of \ac{LoS} \ac{UAV-AP} increases, the \ac{LoS} connection between the user and \ac{UAV-AP} is interrupted, leading to a decrease in \ac{LoS} \ac{UAV} association probability. Moreover, as the height of \ac{LoS} UAV-AP increases, the association probability of \ac{TBS} and \ac{NLoS} UAV-AP increases. The association of \ac{TBS} is more than \ac{NLoS} UAV-AP due to limited blockages during the \ac{TBS} transmission. We can obtain an optimal height for \ac{LoS} \ac{UAV-AP}, so that there is more than 45\% chance for the user to associate to \ac{LoS} UAV-AP or \ac{TBS} and 10\% chance to associate to \ac{NLoS} UAV-AP to be under coverage.  }
On the contrary, Fig.~\ref{fig:x-Haul_asso}, shows that as the intensity of the \acp{TBS} tier increases, there is a monotonic increase and decrease of the \ac{TBS} and \ac{UAV-BS} association for xHaul support at a given \ac{UAV-AP}. 
\vspace{-0.3cm}
\subsection{\ac{SINR} Coverage Probability}

\begin{figure}[!htb]
    \begin{minipage}[t]{0.45\linewidth}
   \centering
    \includegraphics[width=0.9\linewidth]{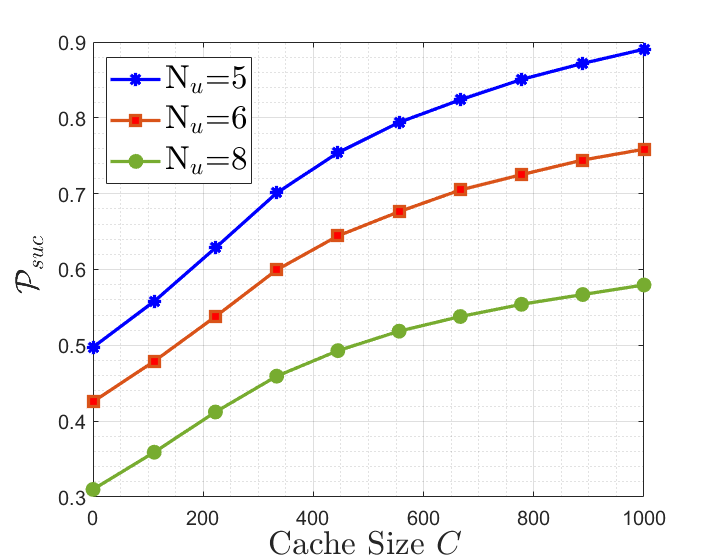}
     \caption{Success probability versus cache size for different values of $N_u$ with $\lambda_{M}=10^{-6} m^{-2}$, $\lambda_{UA}=10^{-5} m^{-3}$, $\eta =9.61$, $\mu =0.16$.}
     \label{fig:success_c}
    \end{minipage}\hfill
    \begin{minipage}[t]{0.45\textwidth}
    \centering
    \includegraphics[width=0.9\linewidth,height=0.7\linewidth]{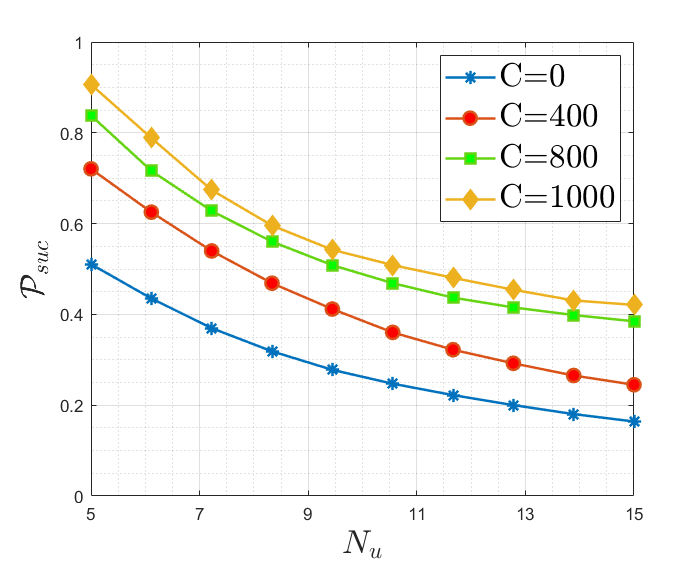}
    \caption{Success Probability versus Number of users in the access link}
    \label{fig:suc_Nu}
    \end{minipage}
\end{figure}

\vspace{-0.2cm}
Fig.~\ref{fig:overall_cov} shows that the analytical result on the overall coverage probability closely matches the Monte-Carlo simulations. We validate the framework for two different values of $\lambda_{UA}$.
Fig.~\ref{fig:success_c} shows the probability of successful content delivery for a different number of simultaneously served users and cache sizes. Naturally, an increase in the cache size or a decrease in the number of users improves the per-user success probability. For the network operator, this reveals how our framework can be used to determine the number of simultaneously served users from one UAV-AP. For example, with $\lambda_{UA}= 10^{-5} m^{-3}$, with a cache size of $C = 600$, the typical user observes about 80\% success if 5 users are served simultaneously. At the same time, it drops to about 50\% if 8 users are served simultaneously. In case the operator wants to sustain success of over 90\%, the operator must necessarily facilitate admission control mechanisms so that no more than 5 users are served simultaneously. This is because with 5 users, even caching all the files in the local storage does not achieve a success probability greater than 0.9.

\begin{figure}[htbp]
\begin{minipage}[t]{0.45\textwidth}
    \centering
    \includegraphics[width=0.9\linewidth]{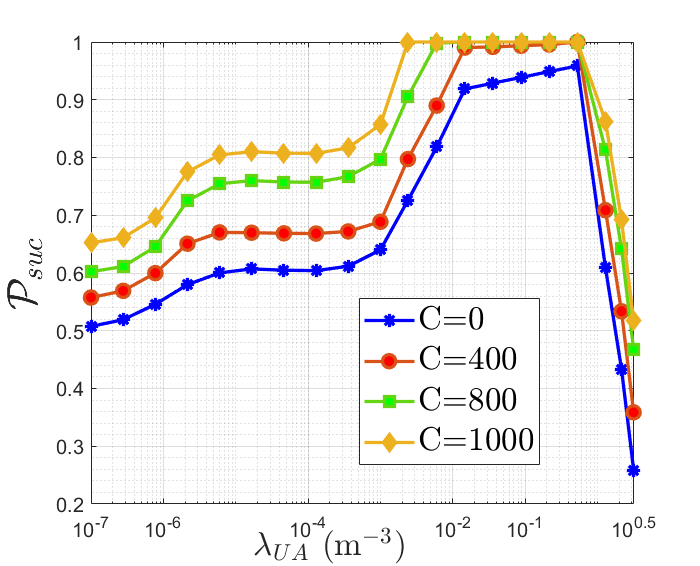}
    \vspace{-0.5cm}
    \caption{Success probability versus intensity of \ac{UAV-AP} for different values of $C$ for $\lambda_{M}=10^{-6} m^{-2}$.}\label{fig:success_intensity}
\end{minipage}\hfill
\begin{minipage}[t]{0.45\textwidth}
   \centering
     \includegraphics[width=0.95\linewidth]{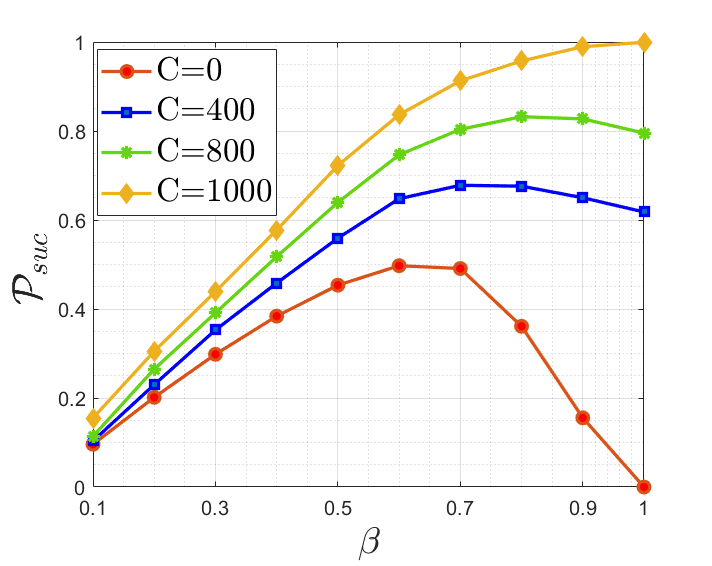}
    \vspace{-0.5cm}
    \caption{Success probability versus the resource partitioning factor for different values of $C$ for $C_0$=$C$/2.}
    \label{fig:success_beta}
\end{minipage}
\end{figure}

\textcolor{blue}{In Fig.~\ref{fig:suc_Nu}, we plot success probability versus the number of users in the access link for different values of cache size. Naturally, as the number of users increases, the success probability in the network decreases. For cache size, $C$=0, gives the lowest success probability as the requirement in the access link increases, and $C$=1000 gives the highest success probability. If the requirement in the access link is increased by three times, the success probability will decrease only by 22\% if we cache more than 400 files at the \ac{UAV-AP} locally. If the files are not cached, the success probability is decreased by 35\%.}
Although, in general the success probability increases with an increase in the number of \acp{UAV-AP}, extreme densification can be detrimental due to increased interference.
In Fig.~\ref{fig:success_intensity} we plot the success probability with respect to $\lambda_{UA}$ for different cache sizes and a fixed resource partitioning factor $\beta = 0.5$. As a dimensioning rule, this prescribes the network operator with deployment densities given the storage capacity of the local cache of the \acp{UAV-AP}. For example, when the \acp{UAV-AP} do not cache any file locally, i.e., $C = 0$, a success probability of beyond 0.9 is obtained only beyond $\lambda_{UA} = 0.01$ m$^{-3}$, On the contrary, with a higher local storage, e.g., $C = 1000$, a success probability of 0.9 is achieved with 10 times fewer \acp{UAV-AP}, i.e., $\lambda_{UA} = 0.001$ m$^{-3}$. In both cases, however, the success probability falls rapidly after $\lambda_{UA} = 0.5$ m$^{-3}$ due to an increase in interference \textcolor{blue}{with densification. This effectively reduces the SINR and rate coverage probability.} Recall from Fig.~\ref{fig:access_asso} that this region corresponds to a higher NLoS association, while all the LoS \acp{UAV-AP} contribute to the interference.
\vspace{-0.1cm}
Until now, we discussed the results with an equal partitioning of frequency resources between the access and the x-hual link. Next, we study the impact of resource partitioning on the success probability.
Fig.~\ref{fig:success_beta} shows the success probability with respect to the resource partitioning factor $\beta$ for different values of cache sizes for the urban scenario. Indeed, $C = 0$ refers to the case when all the files requested by the user from the \ac{UAV-AP} is retrieved from the backhaul connected \ac{UAV-BS}. In this case, $\beta = 1$, i.e., when all the resources are allotted to the access link, results in a 0\% success although the access link achieves a high rate coverage. We note the existence of an optimum value of $\beta$, which maximizes the success probability. When some of the files are stored at the local cache (e.g., $C = 400, 800$, etc.), the operator may provide all the resources to the access link (thereby reducing the xHaul load) without degrading the success probability. In contrast, when all the files are stored in the local cache $C = 1000$, all the frequency resources must necessarily be allotted to the access link. The optimal $\beta$ thus increases with increasing cache size. The optimal $\beta$ trend with its corresponding optimal success probability is discussed next.

\begin{figure}[htbp]
\begin{minipage}[t]{0.45\textwidth}
   \centering
    \includegraphics[width=0.9\linewidth]{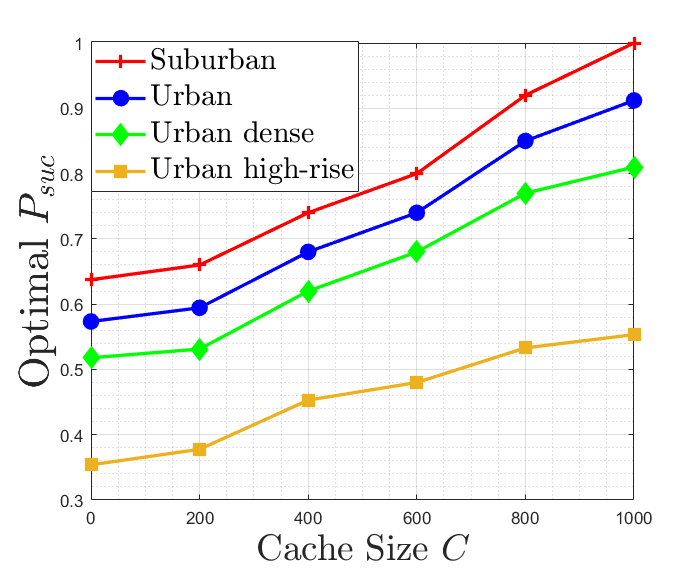}
    \vspace{-0.5cm}
    \caption{Variation of optimal $P_{suc}$ with respect to $C$ for different visibility scenarios}
    \label{fig:optimal_success}
    \end{minipage}\hfill
    \begin{minipage}[t]{0.45\textwidth}
\centering
    \includegraphics[width=0.9\linewidth]{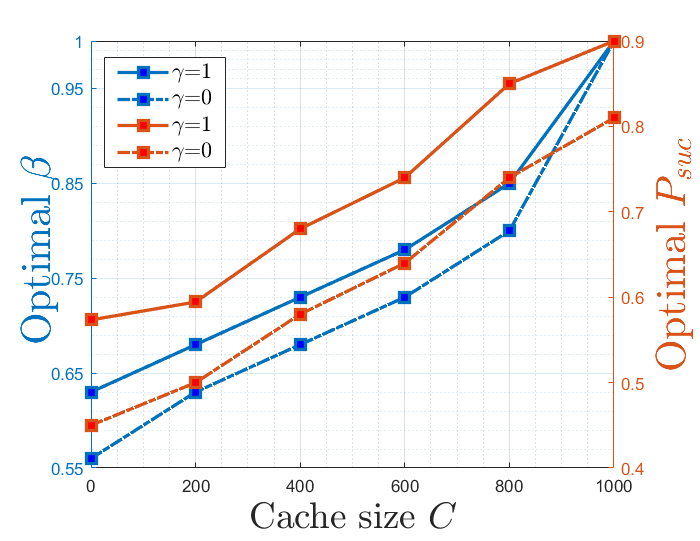}
    \vspace{-0.5cm}
    \caption{Variation of optimal $\beta$  and optimal $P_{suc}$ with respect to $C$ for different values of $\gamma$.}
    \label{fig:optimal_beta}
\end{minipage}
\end{figure}

In Fig.~\ref{fig:optimal_success}, we plot optimal success probability versus cache size for different visibility scenarios. We observe as cache size increases, the probability of successful delivery of contents to the user increases. Let us consider the suburban scenario, where the success probability is high due to less blockages. Remarkably, for $C$=1000, the success in content delivery is 100\% i.e., all the users are delivered with the requested contents directly by the access link. On the contrary, for urban, urban high-rise and dense-urban scenarios, even caching all the files locally at the \acp{UAV-AP} result in lower success.

In Fig.~\ref{fig:optimal_beta}, we plot optimal $\beta$ and success probability versus cache size $C$ for urban scenario. As the cache size increases, optimal $\beta$, along with the maximum success probability increases. In particular, let us consider two cases: $\gamma = 1$ and $\gamma = 0$. For $\gamma = 1$, i.e., when the files are of decreasing popularity, and the \ac{MPC} are cached, the xHaul link is rarely accessed as compared to the case with $\gamma = 0$. Indeed, for $\gamma = 0$, for a given file request from the typical user, a $(1 - C)/1000$ fraction of the time the xHaul support is needed to deliver the file. Accordingly, we observe a higher value of optimal $\beta$, i.e., more resources allocated to the access link for $\gamma = 1$ as compared to $\gamma = 0$. Similarly, due to a more frequent xHaul requirement, the success probability for the case with $\gamma = 0$ is lower as compared to $\gamma = 1$. This reveals that based on the popularity profile of the content, the network operator not only needs to design an optimal access - xHaul split, but also provision minimum local storage at the \acp{UAV-AP}.
For content popularity modeled with $\gamma = 1$, a cache size of $C = 700$ guarantees 85\% success of content delivery. On the contrary, for equiprobable file popularity, even storing all the files at the local cache results in a limited (\~ 80\%) success. In such cases, the operator needs to re-dimension the network with either an increased deployment of \acp{UAV-AP} or provisioning advanced interference management mechanisms. 
For $\gamma=1$, when the files are of decreasing popularity, and the \ac{MPC} files are stored with a probability of one, more bandwidth is given to the access than xHaul. For $\gamma=0$, when the files are of equal popularity, the probability of requesting the files stored in the cache will be less. Therefore, for the successful delivery of files to the user, xHaul is accessed. Thus, the optimal value of $\beta$ is more for $\gamma=1$ compared to $\gamma=0$. Also, for $\gamma=1$, the success probability is more when compared to $\gamma=0$.
\vspace{-0.5cm}
\section{Conclusion}
\label{sec:Con}
In UAV networks, the consideration of the xHaul link capacity and its joint optimization with the access link requirement is imperative. In this work, we derived a joint framework for association and coverage analysis of the access and xHaul links in a UAV-aided cellular network. We showed that for maximizing the LoS association probability, network densification with deploying more UAVs beyond a threshold density does not help, and it deteriorates user performance. Accordingly, we prescribe optimal deployment densities to maximize LoS coverage probability. The distribution of frequency resources among the access and the xHaul link depends on the size of local storage size at the UAV-APs. Larger cache sizes result in a larger allocation of resources in the access link since the xHaul link is used less frequently as compared to smaller cache sizes. We also prescribed admission control strategies in terms of maximum simultaneously served users to sustain a per-user throughput above a threshold. The optimal resource split also depends on relative popularity of the files: the more equi-probable the file popularity is, the larger should be the amount of resources allotted to the xHaul link, especially for small cache sizes. The consideration of mobility of the users and handover between different tiers are interesting directions of research that we will address in future work.
\textcolor{blue}{Also, we will explore different association strategies, other than RSSI-based association scheme, which take performance metrics like throughput directly into account. Moreover, we will investigate spatial stochastic evaluation of optimization frameworks in UAV networks to propose efficient algorithms to optimize network parameters.We will study distributed caching and its impact on system performance and resource partitioning in the future.}

\vspace{-0.5cm}
\section{Acknowledgement}
This research is supported by the IIT Palakkad Technology IHub Foundation Doctoral Fellowship IPTIF/HRD/DF/026.

\begin{appendices}
\vspace{-0.6cm}
\section{Proof of Lemma 2}
The distance distribution to the nearest \ac{LoS} \ac{UAV} is given as:
\begin{gather}
    F_{d_{UAL}}(x)= 1-\mathbb{P}\big(d_{UAL} \textgreater x\big)\nonumber =  1-\exp\Big(-2\pi\lambda_{UA}\int_0^{x} \int_0^{\frac{\pi}{2}} W_L(\theta)y^2\sin{(\theta)}d\theta dy\Big) \nonumber\\
   = 1-\exp\Big(-2\pi\lambda_{UA}\int_0^{x}y^2  W'_L dy\Big)
    \label{app1}
\end{gather}
(\ref{app1}) follows the null probability of 3D PPP. Finally, $f_{d_{UAL}}(x)=\frac{d}{dx}F_{d_{UAL}}(x)$. By following the same steps, $f_{d_{UAN}}(x)$ can also be derived.
\vspace{-0.5cm}
\section{Proof of Lemma 3}
\vspace{-0.5cm}
The CDF of $d_{MB}$ is evaluated as:
\vspace{-0.3cm}
\begin{gather}
 F_{d_{MB}}(x)= 1-\mathbb{P}\big(d_{MB} \textgreater x\big)
 = 1-\exp{\Big(-\pi\lambda_{M}\big(x^2-h^2\big)}\Big)
 \label{app11}
\end{gather}

where $h$ is the height of the typical \ac{UAV-AP} from the ground. Taking the derivation of~(\ref{app11}), we can obtain $f_{d_{MB}}(x)$.
The distance between the \ac{UAV-BS} and \ac{UAV-AP} is $t$.

For $t \leq h$,
\vspace{-0.2cm}
\begin{gather}
   F'_{{d_{UB}}}(x)= 1-\mathbb{P}\big(d_{UB} \textgreater x\big)
   =1-\exp{\Big(-\lambda_{UB}\frac{4}{3}\pi x^3\Big)}
   \label{xHaulpdf}
\end{gather}
\vspace{-0.1cm}
Taking the derivative of (\ref{xHaulpdf}), $f'_{d_{UB}}(x|h)$ is obtained.

For $t > h$, the volume of the segment is given as
\vspace{-0.3cm}
\begin{gather*}
    V_{t-h}=\frac{1}{3}\pi\big(t-h\big)^2\big(3t-(t-h)\big)
    =\frac{1}{3}\pi\big[2t^3-3t^2h+h^3\big]
\end{gather*}

The distance distribution is derived by considering the volume of the sphere from which the volume of the segment $V_{t-h}$ is omitted.
\begin{gather}
    F''_{d_{UB}}(x)=1-\mathbb{P}\big(d_{UB}\textgreater x\big) \nonumber
    =1-\exp{\bigg[-\lambda_{UB}\Big(\frac{4}{3}\pi x^3 - \frac{1}{3}\pi\big[2x^3-3x^2h+h^3\big] \Big)\bigg]} \nonumber\\
    =1-\exp{\bigg[-\lambda_{UB}\Big(\frac{2}{3}\pi x^3 + x^2\pi h - \frac{1}{3}\pi h^3 \Big)\bigg]}
    \label{app22}
\end{gather}
Taking the derivative of (\ref{app22}), we obtain $f''_{d_{UB}}(x|h)$.
\vspace{-0.5cm}
\section{Proof of Lemma 4}
The typical user associates to the \ac{TBS} tier when either of the following events are true:\textcolor{blue}{
Comparing the received powers from all \ac{TBS}, \ac{LoS} \ac{UAV-AP} and \ac{NLoS} \ac{UAV-AP}.}

(i) $R_{MA} > R_{UAL} > R_{UAN}$
(ii) $R_{MA} > R_{UAN} > R_{UAL}$

The probability of event (i) can be written as
\begin{gather}
        \mathbb{P}\Big(P_{M}d_{MA}^{-\alpha_N} > P_{UA}d_{UAL}^{-\alpha_{L}} > P_{UA}d_{UAN}^{-\alpha_{N}}| d_{UAL}\Big)\\
        \textcolor{blue}{\mathbb{P}\Big(d_{UAN}^{\frac{\alpha_{N}}{\alpha_{L}}} > d_{UAL} > C_Md_{MA}^{\frac{\alpha_N}{\alpha_{L}}}\Big)}
        \label{ass1}
\end{gather}
\vspace{-0.3cm}
\textcolor{blue}{Using the cdf of $d_{UAL}$ from $\textbf{Lemma 2}$, for given instances of $d_{UAN}$ and $d_{MA}$, (\ref{ass1}) can be written as:}
\begin{equation}
     \textcolor{blue}{\Big[F_{d_{UAL}}\big(d_{UAN}^{\frac{\alpha_{N}}{\alpha_{L}}}\big) - F_{d_{UAL}}\Big(C_Md_{MA}^{\frac{\alpha_N}{\alpha_{L}}}\Big)|d_{UAN},d_{MA}\Big] \cdot \mathbb{P}\Big(d_{UAN}^{\frac{\alpha_{N}}{\alpha_{L}}} > C_Md_{MA}^{\frac{\alpha_N}{\alpha_{L}}}\Big)} \nonumber
\end{equation}

Considering the cdf terms separately,
\vspace{-0.5cm}
\begin{multline}
F_{d_{UAL}}\left(d_{UAN}^{\frac{\alpha_{N}}{\alpha_{L}}}\right) - F_{d_{UAL}}\left(C_Md_{MA}^{\frac{\alpha_N}{\alpha_{L}}}\right) = \exp{\bigg(-2\pi\lambda_{UA} W'_L \int_0^{\Big(C_Md_{MA}^{\frac{\alpha_N}{\alpha_{L}}}\Big)}z^2d_z\bigg)}-\\ \exp{\bigg(-2\pi\lambda_{UA}W'_L\int_0^{\Big(d_{UAN}^{\frac{\alpha_{N}}{\alpha_{L}}}\Big)}y^2 d_{y}\bigg)}
\label{assacc}
\end{multline}
\vspace{-0.5cm}
\begin{equation}
    \mathbb{P}\bigg(d_{UAN}^{\frac{\alpha_{N}}{\alpha_{L}}} > C_Md_{MA}^{\frac{\alpha_N}{\alpha_{L}}}\bigg)= \mathbb{E}_{d_{MA}} \exp{ \bigg[-2\pi\lambda_{UA} W'_N\int_0^{C_M^{\frac{\alpha_{L}}{\alpha_{N}}}d_M^{\frac{\alpha_N}{\alpha_{N}}}}x^2dx\bigg]}
    \label{assacc1}
\end{equation}
\textcolor{blue}{Taking the expectation with respect to $d_{MA}$ and $d_{UAN}$ over (\ref{assacc}), and combining (\ref{assacc}) and (\ref{assacc1}), we evaluate the probability $A_{MA}^{'}$. }
\begin{multline*}
   \textcolor{blue}{ A'_{MA}= \int_0^\infty \int_{C_{M1}} ^\infty \left[\exp{\left(-\frac{2}{3}\pi\lambda_{UA} W'_L {\bigg(\Big(\frac{P_{UA}}{P_{M}}\Big)^{\frac{3}{\alpha_{L}}}w^{\frac{3\alpha_N}{\alpha_{L}}}\bigg)}\right)}- \exp{\left(-\frac{2}{3}\pi\lambda_{UA}W'_Lx^{\frac{3\alpha_{N}}{\alpha_{L}}}\right)}\right]}\\\textcolor{blue}{f_{d_{UAN}}(x)  dx  \exp{\left(-\frac{2}{3}\pi\lambda_{UA}W'_N\bigg({\Big(\frac{P_{UA}}{P_{M}}\Big)^{\frac{3}{\alpha_{N}}}w^{\frac{3\alpha_N}{\alpha_{N}}}\bigg)}\right)}f_{d_{MA}}(w)dw.}
\end{multline*}
\textcolor{blue}{where $C_{M1}= (\frac{P_{UA}}{P_{M}})^{\frac{1}{\alpha_{N}}} w^{\frac{\alpha_N}{\alpha_{N}}}$}

Similarly, \textcolor{blue}{we can evaluate the probability of event (ii) to obtain $A_{MA}^{''}$, where in the first step, we use the CDF of the variable $d_{UAN}$, and then take the expectation with respect to $d_{MA}$ and $d_{UAL}$.} Finally, adding $A_{MA}^{'}$ and $A_{MA}^{''}$, we can obtain the probability of associating to \ac{TBS} tier, $A_{MA}$.

\textcolor{blue}{Solving with a special case:}
\begin{inparaenum}
    \item \textcolor{blue}{$P_{UA}=P_M$} 
    \item \textcolor{blue}{$\alpha_N=\alpha_L$}
\end{inparaenum}
\begin{equation}
    \textcolor{blue}{A^{'}_{MA} = \frac{W_L^{'}}{W_L^{'} +W_N^{'}}\int_0^\infty \exp{\Bigg(\frac{-2}{3}\pi \lambda_{UA} \Big( W_L^{'} + 2W_{N}^{'}\Big) w^3\Bigg)} f_{d_{MA}}(w) dw}
\end{equation}
\begin{equation}
    \textcolor{blue}{A^{''}_{MA} = \frac{W_N^{'}}{W_L^{'} +W_N^{'}}\int_0^\infty \exp{\Bigg(\frac{-2}{3}\pi \lambda_{UA} \Big( W_N^{'} + 2W_{L}^{'}\Big) w^3\Bigg)} f_{d_{MA}}(w) dw}
\end{equation}

\vspace{-0.5cm}
\section{Proof of Lemma 5}
The typical user associates to the \ac{LoS} \ac{UAV-AP} tier when either of these events are true: \textcolor{blue}{Comparing the received powers from all \ac{TBS}, \ac{LoS} \ac{UAV-AP} and \ac{NLoS} \ac{UAV-AP}.}

(i) $R_{UAL} > R_{UAN} > R_{MA}$
(ii) $R_{UAL} > R_{MA}> R_{UAN}$

\textcolor{blue}{From (\ref{asslos2}), (i) and (ii) can be written as}
\begin{equation}
    \textcolor{blue}{A'_{UAL}(d_{UAL}) = \mathbb{P}\left(R_{UAL} > R_{UAN} > R_{MA}|d_{UAL}\right).}
    \label{a'ul}
\end{equation}
\vspace{-1cm}
\begin{equation}
    \textcolor{blue}{A''_{UAL}(d_{UAL}) = \mathbb{P}\left(R_{UAL} > R_{MA} > R_{UAN}|d_{UAL}\right).}
    \label{a''ul}
\end{equation}
\vspace{-0.3cm}
\textcolor{blue}{(\ref{a'ul}) can be written as,}
\begin{gather}
     \mathbb{P}\Big(P_{UA}d_{UAL}^{-\alpha_{L}} > P_{UA}d_{UAN}^{-\alpha_{N}} > P_{M}d_{MA}^{-\alpha_{N}}\Big)\nonumber
     =\textcolor{blue}{\mathbb{P}\Big(d_{UAL}^{\frac{\alpha_{L}}{\alpha_{N}}}\Big) < d_{UAN} < \Big(\frac{P_{UA}}{P_{M}}\Big)^{\frac{1}{\alpha_{N}}}d_{MA}^{\frac{\alpha_N}{\alpha_{N}}}} \nonumber\\
    \textcolor{blue}{\mathbb{P}\Big(C_{M1}d_{MA}^{\frac{\alpha_N}{\alpha_{N}}} > d_{UAN} > d_{UAL}^{\frac{\alpha_{N}}{\alpha_{L}}}\Big)}
\label{Losass1}
\end{gather}
where $C_{M1}= \big(\frac{P_{UA}}{P_{M}}\big)^{\frac{1}{\alpha_{N}}}$.

\textcolor{blue}{Using cdf of $d_{UAN}$ from $\textbf{Lemma 2}$, for given instances of $d_{UAL}$ and $d_{MA}$, (\ref{Losass1}) written as:}
\begin{equation}
     \textcolor{blue}{\bigg[F_{d_{UAN}}\left(C_{M1}d_{MA}^{\frac{\alpha_N}{\alpha_{N}}}\right)- F_{d_{UAN}}\left(d_{UAL}^{\frac{\alpha_{L}}{\alpha_{N}}}\right)|d_{UAL},d_{MA}\bigg] \cdot \mathbb{P}\left(C_{M1}d_{MA}^{\frac{\alpha_N}{\alpha_{N}}}> d_{UAL}^{\frac{\alpha_{L}}{\alpha_{N}}}\right)} 
\end{equation}
   
\vspace{-1cm}
\begin{multline}
F_{d_{UAN}}\Big(C_{M1}d_{MA}^{\frac{\alpha_N}{\alpha_{N}}}\Big)- F_{d_{UAN}}\Big(d_{UAL}^{\frac{\alpha_{L}}{\alpha_{N}}}\Big) = \exp{\bigg(-2\pi\lambda_{UA} W'_N \int_0^{\Big(d_{UAL}^{\frac{\alpha_{L}}{\alpha_{N}}}\Big)}z^2d_z\bigg)}-\\ \exp{\bigg(-2\pi\lambda_{UA} W'_N\int_0^{\Big(C_{M1}d_{MA}^{\frac{\alpha_N}{\alpha_{N}}}\Big)}y^2 d_{y}\bigg)} 
\label{assnlos1}
\end{multline}
\begin{equation}
    \mathbb{P}\bigg(d_{MA} > \frac{1}{C_{M1}}d_{UAL}^{\frac{\alpha_{L}}{\alpha_N}}\bigg)= \exp{\Big(-\pi\lambda_M \cdot \Big(\Big(\frac{1}{C_{M1}}\Big)^{\frac{\alpha_{N}}{\alpha_N}}d_{UAL}^{\frac{\alpha_{L}}{\alpha_N}}\Big)^2}\Big)
    \label{assnlos2}
\end{equation}
\textcolor{blue}{Taking the expectation wrt $d_{MA}$ over (\ref{assnlos1}), and combining (\ref{assnlos1}) and (\ref{assnlos2}), we can obtain $A'_{UAL}(d_{UAL})$.}
\begin{multline*}
   \textcolor{blue}{ A'_{UAL}(d_{UAL}) = \int_{C_{L1}}^\infty \bigg[\exp{\Big(-\frac{2}{3}\pi\lambda_{UA} W'_N d_{UAL}^{\frac{3\alpha_{L}}{\alpha_{N}}}\Big)}- \exp\Big(-\frac{2}{3}\pi\lambda_{UA}W'_N(\frac{P_{UA}}{P_{M}})^{\frac{3}{\alpha_{N}}}x^{\frac{3\alpha_N}{\alpha_{N}}}\Big)\bigg]}\\\textcolor{blue}{ \exp{\Big(-\pi\lambda_{M} {\left(\frac{P_{M}}{P_{UA}}\right)^{\frac{2}{\alpha_N}}d_{UAL}^{\frac{2\alpha_{L}}{\alpha_N}}}\Big)}f_{d_{MA}}(x) dx.}
\end{multline*}
where $C_{L1}= {(\frac{P_{M}}{P_{UA}})^{\frac{1}{\alpha_N}}d_{UAL}^{\frac{\alpha_{L}}{\alpha_N}}},  L_1=(\frac{P_{UA}}{P_{M}})^{\frac{1}{\alpha_{N}}}x^{\frac{\alpha_N}{\alpha_{N}}}$.

\textcolor{blue}{Similarly, we can evaluate the probability of event (ii) to obtain $A''_{UAL}(d_{UAL})$ from (\ref{a''ul}), where in the first step, we use the CDF of $d_{MA}$ and then take the expectation with respect to $d_{UAN}$. Finally, combining $A_{UAL}^{'}$ and $A_{UAL}^{''}$, and taking the expectation with respect to $d_{UAL}$, we can obtain the probability of associating to \ac{LoS} \ac{UAV-AP} tier, $\bar{A}_{UAL}$, }given by $\bar{A}_{UAL} =    \int_0^{\infty} A_{UAL}(x) f_{d_{UAL}}(x) dx 
    \label{a_bar_los}$.
\vspace{-1cm}
\section{Proof of Proposition 1}
\textcolor{blue}{Recall the probability of \ac{LoS} association in access link $\bar{A}_{UAL}$ is divided into two parts as in (\ref{asslos2}) and (\ref{a_bar_los}) as:}
\begin{equation}
    \bar{A}_{UAL}= \int_0^\infty A'_{UAL}(d_{UAL}) f_{d_{UAL}}(x) dx + \int_0^\infty A^{''}_{UAL}(d_{UAL}) f_{d_{UAL}}(x) dx 
    \label{losassn123}
\end{equation}

The first part can be written as
\begin{multline}
    \int_0^\infty A'_{UAL}(d_{UAL}) f_{d_{UAL}}(x) dx  = \int_0^\infty \int_{C_{L1}}^\infty \bigg[\exp\bigg(-2\pi\lambda_{UA} W'_N \int_0^{d_{UAL}^{\frac{\alpha_{L}}{\alpha_{N}}}}z^2d_z \bigg) \\ -\exp\Big(-2\pi\lambda_{UA}W'_N\int_0^{L_1}y^2 d_{y}\Big)\bigg] \cdot \exp{\Big(-\pi\lambda_{M} (C_{L1})^2\Big)} f_{d_{MA}}(x) f_{d_{UAL}}(y) dx dy \nonumber
\end{multline}
\vspace{-1cm}
\begin{multline}
    =\textcolor{blue}{\underbrace{\int_0^\infty \int_{C_{L1}}^\infty \exp\bigg(-2\pi\lambda_{UA} W'_N \int_0^{d_{UAL}^{\frac{\alpha_{L}}{\alpha_{N}}}}z^2d_z \bigg) \exp{\Big(-\pi\lambda_{M} (C_{L1})^2\Big)} f_{d_{MA}}(x) f_{d_{UAL}}(y) dx dy}_{\RomanNumeralCaps{1}} -} \\ \textcolor{blue}{\underbrace{\int_0^\infty \int_{C_{L1}}^\infty \exp\Big(-2\pi\lambda_{UA}W'_N\int_0^{L_1}y^2 d_{y}\Big)\exp{\Big(-\pi\lambda_{M} (C_{L1})^2\Big)} f_{d_{MA}}(x) f_{d_{UAL}}(y) dx dy}_{\RomanNumeralCaps{2}}}
    \label{losproposition}
\end{multline}
where $C_{L1}= {(\frac{P_{M}}{P_{UA}})^{\frac{1}{\alpha_N}}d_{UAL}^{\frac{\alpha_{L}}{\alpha_N}}}$ , $L_1=(\frac{P_{UA}}{P_{M}})^{\frac{1}{\alpha_{N}}}d_{MA}^{\frac{\alpha_N}{\alpha_{N}}}$. 
To prove that there exists at least one maxima of the association probability with respect to $\lambda_{UA}$, let us observe the derivative of the first term  of (\ref{losproposition}) that constitutes $A'_{UAL}$:
\begin{multline}
 \frac{d}{d\lambda_{UA}}\Bigg[\int_0^\infty \int_{C_{L1}}^\infty \exp\Big(-\frac{2}{3}\pi \lambda_{UA}W'_Ny^{\frac{3\alpha_L}{\alpha_N}}\Big)\exp\Big(-\pi \lambda_{M}\Big[\Big(\frac{P_M}{P_{UA}}\Big)^{\frac{2}{\alpha_N}}y^{\frac{2\alpha_L}{\alpha_N}}\Big]\Big)\\\cdot 2\pi \lambda_M x\exp{\big(-\pi\lambda_M x^2\big)} 2\pi \lambda_{UA}W'_L y^2 \exp\Big(-\frac{2}{3}\pi\lambda_{UA}W'_Ly^3\Big) dx dy \Bigg] \nonumber
\end{multline}
\begin{multline}
    \frac{d}{d\lambda_{UA}}\Bigg[\int_0^\infty 2\pi \lambda_{UA}y^2W'_L \exp\bigg(-2\pi\lambda_{M} \Big[\Big(\frac{P_{M}}{P_{UA}}\Big)^{\frac{2}{\alpha_N}}y^{\frac{2\alpha_L}{\alpha_N}}\Big]\bigg) \exp\Big(-\frac{2}{3}\pi\lambda_{UA}\Big(W'_Ny^{\frac{3\alpha_L}{\alpha_N}}+W'_Ly^3\Big)\Big) dy\Bigg] 
    \label{maxlos}
\end{multline}
Applying Leibniz Integral rule in (\ref{maxlos}),
\begin{multline}
 \int_0^\infty \Bigg[ -\frac{4}{3}\pi \lambda_{UA}\Big(W'_Ny^{\frac{3\alpha_L}{\alpha_N}}+W'_Ly^3\Big) y^2W'_L\exp\bigg(-2\pi\lambda_{M}\Big[\Big(\frac{P_{M}}{P_{UA}}\Big)^{\frac{2}{\alpha_N}}y^{\frac{2\alpha_L}{\alpha_N}}\Big]\bigg)\\ \exp\bigg(-\frac{2}{3}\pi\lambda_{UA}\Big(W'_Ny^{\frac{3\alpha_L}{\alpha_N}}+W'_Ly^3\Big)\bigg) + 2\pi y^2W'_L\exp\bigg(-2\pi\lambda_M \Big[\Big(\frac{P_{M}}{P_{UA}}\Big)^{\frac{2}{\alpha_N}}y^{\frac{2\alpha_L}{\alpha_N}}\Big]\bigg) \\ \exp\bigg(-\frac{2}{3}\pi\lambda_{UA}\Big(W'_Ny^{\frac{3\alpha_L}{\alpha_N}}+W'_Ly^3\Big)\bigg)\Bigg]
 \label{assprop4}
\end{multline}

At $\lambda_{UA}=0$, (\ref{assprop4}) becomes $\int_0^\infty 2\pi y^2W'_L\exp{\bigg(-2 \pi \lambda_M\Big[\Big(\frac{P_{M}}{P_{UA}}\Big)^{\frac{2}{\alpha_N}}y^{\frac{2\alpha_L}{\alpha_N}}\Big]\bigg)}dy$,

\textcolor{blue}{which gives a positive value when substituting the values and integrate with respect to y.} On the other hand, the derivative of ${\RomanNumeralCaps{2}}$ of (\ref{losproposition}) constitutes $A'_{UAL}$.
Therefore, the derivative of $A'_{UAL}$ is when $\lambda_{UA}$ is zero. \textcolor{blue}{Similarly, we can prove the derivative of $\int_0^{\infty}A''_{UAL}(d_{UAL}) f_{d_{UAL}}(x) dx$ in (\ref{losassn123}), with respect to $\lambda_{UA}$ is also a positive function when $\lambda_{UA}=0$. Thus, at $\lambda_{UA} = 0$, $\bar{A}_{UAL}$ is an increasing function of $\lambda_{UA}$.}
\textcolor{blue}{On the contrary, substituting $\lambda_{UA}= \infty$ directly in the expansion of (\ref{losassn123}), we note that both evaluate to zero.
Consequently, there exists at least one maxima of $\bar{A}_{UAL}$ with respect to $\lambda_{UA}$, as $\bar{A}_{UAL}$ is an increasing function for $\lambda_{UA}$=0, and $\bar{A}_{UAL}$ is zero when $\lambda_{UA}$= $\infty$. Therefore, we can say that, there exists optimal UAV densities which maximizes the probability of association of typical user with the \ac{LoS} \ac{UAV-AP}.}


\vspace{-0.8cm}

\section{Proof of Lemma 7}
Given an access link distance of $d_a$, the \ac{UAV-AP} associates to a \ac{UAV-BS} in case the received power from it is larger than the one received from a \ac{TBS}. This probability is evaluated as:
\begin{equation}
        A_{UB}(d_a, \theta) =  \mathbb{P}\left(P_{UB}d_{UB}^{-\alpha_L} \textgreater P_{M}d_{MB}^{-\alpha_L} | d_a\right) = \mathbb{P}\left(d_{UB} \leq \left(\frac{P_M}{P_{UB}} d^{-\alpha_L}_{MB}\right)^{-\frac{1}{\alpha_L}} | d_a \right) = \mathbb{E}_{d_{MB}}\left[\mathcal{T}(d_{MB})\right] \nonumber 
\end{equation}
        where,
        \begin{align}
        &\mathcal{T}(d_{MB}) = &\begin{cases}
         1 - \exp\left(-\frac{4}{3}\pi \lambda_{UB} \left(\frac{P_M}{P_{UB}} d_{MB}^{-\alpha_L}\right)^{-\frac{3}{\alpha_L}} \right); 
        \qquad
        d_{MB}  \leq \left(\left(d_a \cos(\theta)\right)^{-\alpha_L}\frac{P_{UB}}{P_M}\right)^{-\frac{1}{\alpha_L}} \nonumber\\
        F''_{d_{UB}}\left(\left(\frac{P_M}{P_{UB}} d_{MB}^{-\alpha_L}\right)^{-\frac{1}{\alpha_L}}\right);
         \qquad d_{MB}  > \left(\left(d_a \cos(\theta)\right)^{-\alpha_L}\frac{P_{UB}}{P_M}\right)^{-\frac{1}{\alpha_L}}
        \end{cases}
\end{align}
The expectation is taken with respect to $f_{d_{MB}}$, which is defined only for $x \geq h$. Considering,
\vspace{-0.3cm}
\begin{align}
    \ell(d_a, \theta) =  \left(\left(d_a \cos(\theta)\right)^{-\alpha_L}\frac{P_{UB}}{P_M}\right)^{-\frac{1}{\alpha_L}} \nonumber
\end{align}
we note that for $P_{UB} \leq P_M$, we have $\left(\frac{P_{UB}}{P_M}\right)^{-\frac{1}{\alpha_L}} \geq 1$ and accordingly, $\left(\left(d_a \cos(\theta)\right)^{-\alpha_L}\frac{P_{UB}}{P_M}\right)^{-\frac{1}{\alpha_L}} \geq d \cos(\theta)$. Accordingly, the expectation with respect to $d_{MB}$ evaluates to:
\begin{multline}
    A_{UB}(d_a, \theta) = \int_{d_a \cos(\theta)}^{\ell(d_a, \theta)} 1 - \exp\left(-\frac{4}{3}\pi \lambda_{UB} \left(\frac{P_M}{P_{UB}} x^{-\alpha_L}\right)^{-\frac{3}{\alpha_L}}\right) f_{d_{MB}}(x)dx  + \\
     \int_{\ell(d_a, \theta)}^\infty F''_{d_{UB}}\left(\left(\frac{P_M}{P_{UB}} d_{MB}^{-\alpha_L}\right)^{-\frac{1}{\alpha_L}}\right) f_{d_{MB}}(x)dx
\end{multline}
\vspace{-1.5cm}
\section{Proof of Lemma 8}
Given that the typical user has associated to \ac{TBS}, the \ac{SINR} coverage probability is given as
\begin{equation}
    \mathcal{P}_{CM}=     \mathbb{P}\Bigg[\frac{K_MP_{M}d_{MA}^{-\alpha_N}g_M}{N_0+I_{M1}+I_{L1}+I_{NL1}} \textgreater t_a \Bigg]
    \label{sinr1}
\end{equation}
where $I_{M1}$, $I_{L1}$ and $I_{NL1}$ are the  interference strength from the tier of \acp{TBS}, \ac{LoS} \acp{UAV-AP} and \ac{NLoS} \acp{UAV-BS} respectively, where $I_{M1}= \sum_{l:X_l \in \Phi^{'}_{M}}K_MP_{M}d'^{-\alpha_N}_{M,l}\bar{g}'_{l}$, $I_{L1}= \sum_{l:X_l \in \Phi_L}K_UP_{UA}d_{L,l}^{-\alpha_{L}}\bar{G}_l$ and $I_{NL1}= \sum_{l:X_l \in \Phi_{N}}K_UP_{UA}d_{N,l}^{-\alpha_{N}}\hat{g}_{l}$. $\Phi^{'}_{M}$ is the tier of \acp{TBS} in which the associated \ac{TBS} is omitted. $d'_{M}$ is the distance of user from the \acp{TBS} other than the associated \ac{TBS}.  $d_{L}$ is the distance of user from the interfering LoS \acp{UAV-AP}. $d_{N}$ is the distance of user from the interfering NLoS \acp{UAV-AP}. $\bar{g}$, $\bar{G}$ and $\hat{g}$ are the fast-fading coefficients from the interfering \acp{TBS}, \ac{LoS} \acp{UAV-AP} and \ac{NLoS} \acp{UAV-AP} respectively.
\vspace{-0.1cm}
Taking the expectation over the individually independent \ac{TBS}, LoS/NLoS \ac{UAV-AP} tiers in (\ref{sinr1}), the interference terms are expressed as,
\begin{gather*}
    I'_{M1}= \mathbb{E}_{\Phi^{'}_{M},\bar{g'}} \bigg[\exp{\left(\frac{-t_a I_{M1}}{{K_MP_{M}d_{MA}^{-\alpha_N}}}\right)}\bigg]
    = \mathbb{E}_{\Phi'_{M}}\bigg[\prod_{l:X_l \in \Phi'_M} \frac{1}{1+\frac{t_a d'^{-\alpha_N}_{M,l}}{d_{MA}^{-\alpha_N}}}\bigg]
\end{gather*}
Computing the moment generating function of exponential random variable $\bar{g}$, (\ref{intertbs}) is obtained. 
\vspace{-0.1cm}
\begin{gather}
    I'_{L1}=\mathbb{E}_{\Phi_L,\bar{G}}\bigg[\exp{\left(\frac{- t_a I_{L1}}{{K_MP_{M}d_{MA}^{-\alpha_N}}} \right)}\bigg]
    =\mathbb{E}_{\Phi_L}\Bigg[\prod_{l:X_l \in \Phi_L} \Bigg(\frac{m}{m+\frac{\eta t_a K_UP_{UA}d_{L,l}^{-\alpha_{L}}}{{{K_MP_{M}d_{MA}^{-\alpha_N}}}}}\Bigg)^m\Bigg] 
    \label{sinrinterlos}
\end{gather}
where $\eta= m(m!)^{\frac{-1}{m}}$. Noting that $|\bar{G}|^2$ is a normalized gamma random variable with parameter $m$. Computing moment generating function of gamma random variable $\bar{G}$, we can obtain (\ref{sinrlosinterference}).
\vspace{-0.3cm}
\begin{gather}
    I'_{NL1}=\mathbb{E}_{\Phi_{N},\hat{g}} \bigg[\exp{\left(\frac{-t_a I_{NL1}}{{K_MP_{M}d_{MA}^{-\alpha_N}}}\right)}\bigg] 
    = \mathbb{E}_{\Phi_{N}}\bigg[\prod_{l:X_l \in \Phi_N}\frac{1}{1+\frac{t_a K_UP_{UA}d_{N,l}^{-\alpha_{N}}}{{K_MP_{M}d_{MA}^{-\alpha_N}}}}\bigg] \nonumber
\end{gather}
Solving these equations and substituting in (\ref{sinrmbs}), we can obtain $\mathcal{P}_{CM}$.

Similarly, the \ac{SINR} coverage probability of typical user associated to \ac{LoS} UAV is given as
\begin{equation}
    \mathcal{P}_{CL}=\mathbb{P}\Bigg[\frac{K_UP_{UA}d_{UAL}^{-\alpha_{L}}G_L}{N_0+I_{M3}+I_{L3}+I_{NL3}} \textgreater t_a \Bigg]
    \label{sinr2}
\end{equation}
where $I_{M3}$, $I_{L3}$ and $I_{NL3}$ are the  interference strength from the tier of \acp{TBS}, \ac{LoS} \acp{UAV-AP} and \ac{NLoS} \acp{UAV-BS} respectively,  where $I_{M3}=\sum_{l:X_l \in \Phi_{M}}K_MP_{M}d_{M,l}^{-\alpha_N}\bar{g}_l$, $I_{L3}=\sum_{l:X_l \in \Phi^{'}_{L}}K_UP_{UA}d'^{-\alpha_{L}}_{L,l}\bar{G}^{'}_l$ and $I_{NL3}= \sum_{l:X_l \in \Phi_{N}}K_UP_{UA}d_{N,l}^{-\alpha_{N}}\hat{g}_{l}$. $\Phi^{'}_{L}$ is the tier of LoS \acp{UAV-AP} in which the associated LoS \ac{UAV-AP} is omitted. $d'_{L}$ is the distance of user from the LoS \acp{UAV-AP} other than the associated LoS \ac{UAV-AP}. $\bar{G}^{'}$ is the fast-fading coefficient from the interfering LoS \acp{UAV-AP} other than the associated LoS \ac{UAV-AP}.

 (\ref{sinr2}) can be written as,
\begin{equation*}
    \mathcal{P}_{CL}= 1-\mathbb{E}_{\Phi}\bigg[\left(1- \exp{\left(\frac{-\eta t_a (N_0 + I_{M3} +I_{NL3} + I_{L3})}{K_UP_{UA}d_{UAL}^{-\alpha_{L}}} \right)}\right)^m\bigg]
    \end{equation*}
where $\Phi$ is the union of the individual independent \ac{PPP}. $\Phi= \Phi_M \cup \Phi'_L \cup \Phi_N$.

The above equation following the Binomial theorem can be written as,
    \begin{equation}
    \sum_{n=1}^m (-1)^{n+1} \Comb{m}{n}  \mathbb{E}_{\Phi} \bigg[\exp{\left(\frac{-n \eta t_a \big(N_0 + I_{M3} +I_{NL3} + I_{L3}\big)}{K_UP_{UA}d_{UAL}^{-\alpha_{L}}} \right)}\bigg]
    \label{inter11}
\end{equation}
 Therefore, by applying expectation over the tiers in (\ref{inter11}), the interference terms are expressed as,
 \vspace{-0.8cm}
\begin{gather}
    I'_{M3}= \mathbb{E}_{\Phi_{M},\bar{g}}\bigg[\exp{\left(\frac{-n \eta t_a I_{M3}}{{K_UP_{UA}d_{UAL}^{-\alpha_{L}}}}\right)}\bigg] \nonumber,
    I'_{L3}= \mathbb{E}_{\Phi^{'}_{L},\bar{G}^{'}}\bigg[\exp{\left(\frac{-n \eta t_a I_{L3}}{{K_UP_{UA}d_{UAL}^{-\alpha_{L}}}} \right)}\bigg] \nonumber\\
    = \mathbb{E}_{\Phi^{'}_{L}}\bigg[\prod_{l:X_l \in \Phi'_L}  \mathbb{E}_{\bar{G}^{'}}\left(\exp{\left(\frac{-n \eta t_a K_UP_{UA}d'^{-\alpha_{L}}_{L,l}\bar{G}^{'}_l}{{K_UP_{UA}d_{UAL}^{-\alpha_{L}}}} \right)}\right)\bigg]
    =\mathbb{E}_{\Phi_{L'}}\Bigg[\prod_{l:X_l \in \Phi'_L} \left(\frac{m}{m+\frac{n \eta t_a d'^{-\alpha_{L}}_{L,l}}{{{d_{UAL}^{-\alpha_{L}}}}}}\right)^m\Bigg] \nonumber
\end{gather}
Computing the moment generating function of gamma random variable $\bar{G}$, we can obtain (\ref{interflos}).
\vspace{-0.3cm}
\begin{gather}
    I'_{NL3}= \mathbb{E}_{\Phi_{N},\hat{g}}\bigg[\exp{\left(\frac{-n \eta t_a I_{NL3}}{{K_UP_{UA}d_{UAL}^{-\alpha_{L}}}}\right)}\bigg] \nonumber
\end{gather}
Solving these equations and substituting in (\ref{sinrlos}), we can obtain $P_{CL}$. 
Similarly, we can obtain the \ac{SINR} coverage probability of typical user associated to \ac{NLoS} UAV, $\mathcal{P}_{CN}$ given in (\ref{sinrnlos}).

The proof of \ac{SINR} coverage probability of tagged \ac{UAV-AP} associated to \ac{TBS} or \ac{UAV-BS} in the xHaul link follows in a similar way as Proof of Lemma 8.

\end{appendices}
\vspace{-0.3cm}
\bibliography{references}

\begin{thebibliography}{10}
\providecommand{\url}[1]{#1}
\csname url@samestyle\endcsname
\providecommand{\newblock}{\relax}
\providecommand{\bibinfo}[2]{#2}
\providecommand{\BIBentrySTDinterwordspacing}{\spaceskip=0pt\relax}
\providecommand{\BIBentryALTinterwordstretchfactor}{4}
\providecommand{\BIBentryALTinterwordspacing}{\spaceskip=\fontdimen2\font plus
\BIBentryALTinterwordstretchfactor\fontdimen3\font minus
  \fontdimen4\font\relax}
\providecommand{\BIBforeignlanguage}[2]{{%
\expandafter\ifx\csname l@#1\endcsname\relax
\typeout{** WARNING: IEEEtran.bst: No hyphenation pattern has been}%
\typeout{** loaded for the language `#1'. Using the pattern for}%
\typeout{** the default language instead.}%
\else
\language=\csname l@#1\endcsname
\fi
#2}}
\providecommand{\BIBdecl}{\relax}
\BIBdecl

\bibitem{16}
A.~M. Hayajneh, S.~A.~R. Zaidi, D.~C. McLernon, and M.~Ghogho, ``Drone
  empowered small cellular disaster recovery networks for resilient smart
  cities,'' \emph{2016 IEEE international conference on sensing, communication
  and networking (SECON Workshops)}, pp. 1--6, Jun 2016.

\bibitem{17}
M.~Hua, Y.~Wang, Z.~Zhang, C.~Li, Y.~Huang, and L.~Yang, ``Power-efficient
  communication in {UAV}-aided wireless sensor networks,'' \emph{IEEE
  Communications Letters 22.6}, pp. 1264--1267, Apr 2018.

\bibitem{18}
P.~Ji, X.~Jia, Y.~Lu, H.~Hu, and Y.~Ouyang, ``Multi-{UAV} assisted multi-tier
  millimeter-wave cellular networks for hotspots with {2-Tier} and {4-Tier}
  network association,'' \emph{IEEE Access 8}, pp. 158\,972--158\,995, Aug
  2020.

\bibitem{35}
D.~Harutyunyan and R.~Riggio, ``Flex{5G}: Flexible functional split in {5G}
  networks,'' \emph{IEEE Transactions on Network and Service Management},
  vol.~15, no.~3, pp. 961--975, 2018.

\bibitem{21}
M.~Mozaffari, W.~Saad, M.~Bennis, Y.-H. Nam, and m.~Debbah, ``A tutorial on
  {UAV}s for wireless networks: Applications, challenges, and open problems,''
  pp. 2334--2360, 03 2018.

\bibitem{22}
M.~Mozaffari, W.~Saad, M.~Bennis, and M.~Debbah, ``Mobile unmanned aerial
  vehicles {(UAVs)} for energy-efficient internet of things communications,''
  \emph{IEEE Transactions on Wireless Communications 16}, pp. 7574--7589, Sep
  2017.

\bibitem{23}
Bor-Yaliniz, R.~Irem, A.~El-Keyi, and H.~Yanikomeroglu, ``Efficient {3-D}
  placement of an aerial base station in next generation cellular networks,''
  \emph{016 IEEE international conference on communications (ICC)}, pp. 1--5,
  May 2016.

\bibitem{hetnets}
Y.~J. Chun, M.~O. Hasna, and A.~Ghrayeb, ``Modeling heterogeneous cellular
  networks interference using poisson cluster processes,'' \emph{IEEE Journal
  on Selected Areas in Communications}, vol.~33, pp. 1--1, 10 2015.

\bibitem{cache}
F.~Cheng, G.~Gui, N.~Zhao, Y.~Chen, J.~Tang, and H.~Sari,
  ``{UAV}-relaying-assisted secure transmission with caching,'' \emph{IEEE
  Transactions on Communications}, vol.~67, pp. 3140--3153, 5 2019.

\bibitem{24}
A.~Fotouhi, H.~Qiang, M.~Ding, M.~Hassan, L.~G. Giordano, A.~Garcia-Rodriguez,
  and J.~Yuan, ``Survey on {UAV} cellular communications: Practical aspects,
  standardization advancements, regulation, and security challenges,''
  \emph{IEEE Communications Surveys \& Tutorials}, vol. 21.4, pp. 3417--3442,
  March 2019.

\bibitem{25}
A.~Al-Hourani, S.~Kandeepan, and S.~Lardner, ``Optimal {LAP} altitude for
  maximum coverage,'' \emph{IEEE Wireless Communications Letters 3.6}, pp.
  569--572, Jul 2014.

\bibitem{32}
D.~C. Chen, T.~Q. Quek, and M.~Kountouris, ``Wireless backhaul in small cell
  networks: Modelling and analysis,'' \emph{2014 IEEE 79th Vehicular Technology
  Conference (VTC Spring). IEEE, 2014.}, pp. 1--6, May 2014.

\bibitem{33}
N.~Tafintsev, D.~Moltchanov, M.~Gerasimenko, M.~Gapeyenko, j.~Zhu, S.-p. Yeh,
  N.~Himayat, S.~Andreev, Y.~Koucheryavy, and M.~Valkama, ``Aerial access and
  backhaul in mmwave b5g systems: Performance dynamics and optimization,''
  \emph{IEEE Communications Magazine 58.2}, pp. 93--99, Feb 2020.

\bibitem{26}
e.~a. Polese, Michele, ``Integrated access and backhaul in {5G} mmwave
  networks: Potential and challenges,'' \emph{IEEE Communications Magazine
  58.3}, pp. 62--68, Mar 2020.

\bibitem{8}
C.~Pan, J.~Yi, C.~Yin, J.~Yu, and X.~Li, ``Joint {3D} {UAV} placement and
  resource allocation in software-defined cellular networks with wireless
  backhaul,'' \emph{IEEE Access, 7}, pp. 104\,279--104\,293, July 2019.

\bibitem{27}
E.~Kalantari, M.~Z. Shakir, and A.~Yongacoglu, ``Backhaul-aware robust {3D}
  drone placement in {5G}+ wireless networks,'' pp. 109--114, 05 2017.

\bibitem{28}
E.~Kalantari, H.~Yanikomeroglu, and A.~Yongacoglu, ``Wireless networks with
  cache-enabled and backhaul-limited aerial base stations,'' \emph{IEEE
  Transactions on Wireless Communications 19.11}, pp. 7363--7376, July 2020.

\bibitem{7}
A.~A. Khuwaja, Y.~Zhu, G.~Zheng, Y.~Chen, and W.~Liu, ``Performance analysis of
  hybrid {UAV} networks for probabilistic content caching,'' \emph{IEEE Systems
  Journal}, Aug 17 2020.

\bibitem{29}
T.~Zhang, Z.~Wang, Y.~Liu, W.~Xu, and A.~Nallanathan, ``Caching placement and
  resource allocation for cache-enabling {UAV} {NOMA} networks,'' \emph{IEEE
  Transactions on Vehicular Technology, 69.11}, pp. 12\,897--12\,911, Aug 2020.

\bibitem{30}
------, ``Cache-enabling {UAV} communications: Network deployment and resource
  allocation,'' \emph{IEEE Transactions on Wireless Communications 19.11}, pp.
  7470--7483, Jul 2020.

\bibitem{10}
J.~Rao, H.~Feng, C.~Yang, Z.~Chen, and B.~Xia, ``Optimal caching placement for
  {D2D} assisted wireless caching networks,'' \emph{2016 IEEE international
  conference on communications (ICC)}, pp. 1--6, May 2016.

\bibitem{13}
A.~Fouda, A.~S. Ibrahim, I.~Guvenc, and M.~Ghosh, ``{UAV}-based in-band
  integrated access and backhaul for {5G} communications,'' \emph{2018 IEEE
  88th Vehicular Technology Conference (VTC-Fall)}, pp. 1--5, Aug 2018.

\bibitem{31}
E.~Baştuǧ, B.~Mehdi, M.~Kountouris, and M.~Debbah, ``Cache-enabled small cell
  networks: Modeling and tradeoffs,'' \emph{EURASIP Journal on Wireless
  Communications and Networking 2015.1}, pp. 1--11, Dec 2015.

\bibitem{3}
H.~Sun, X.~Wang, C.~Xu, Y.~Zhang, and T.~QS~Quek, ``Performance analysis for
  drone-assisted hetnets with flexible cell association,'' \emph{ICC 2020-2020
  IEEE International Conference on Communications (ICC)}, pp. 1-6 2020.

\bibitem{11}
T.~Bai and R.~W.~Heath, ``Coverage and rate analysis for millimeter-wave
  cellular networks,'' \emph{IEEE Transactions on Wireless Communications 14},
  pp. 1100--14, Oct 2014.

\bibitem{6}
B.~Galkin, J.~Kibiłda, and L.~A. DaSilva, ``A stochastic geometry model of
  backhaul and user coverage in urban {UAV} networks,'' \emph{arXiv preprint
  arXiv:1710.03701}, Oct 9 2017.

\bibitem{34}
I.~Atzeni, J.~Arnau, and M.~Kountouris, ``Downlink cellular network analysis
  with los/nlos propagation and elevated base stations,'' \emph{IEEE
  Transactions on Wireless Communications 17.1}, pp. 142--156, Oct 2017.

\bibitem{4}
Y.~Zhu, G.~Zheng, L.~Wang, K.-K. Wong, and L.~Zhao, ``Content placement in
  cache-enabled {sub-6 {GHz}} and millimeter-wave multi-antenna dense small
  cell networks,'' \emph{IEEE Transactions on Wireless Communications 17, no.
  5}, pp. 2843--2856, 2018.

\bibitem{14}
J.~Qiao, Y.~He, and X.~S. Shen, ``Proactive caching for mobile video streaming
  in millimeter wave {5G} networks,'' \emph{IEEE Transactions on Wireless
  Communications}, pp. 7187--98, Aug 2016.

\bibitem{5}
X.~Lin, J.~Xia, and Z.~Wang, ``Probabilistic caching placement in
  {UAV}-assisted heterogeneous wireless networks,'' \emph{Physical
  Communication 33}, pp. 54--61, 2019.

\bibitem{chiu2013stochastic}
S.~N. Chiu, D.~Stoyan, W.~S. Kendall, and J.~Mecke, \emph{Stochastic geometry
  and its applications}.\hskip 1em plus 0.5em minus 0.4em\relax John Wiley \&
  Sons, 2013.

\bibitem{38}
W.~Yi, Y.~Liu, , and A.~Nallanathan, ``Modeling and analysis of {D2D}
  millimeter-wave networks with poisson cluster processes,'' \emph{IEEE
  Transactions on Communications}, pp. 5574--5588, Aug 2017.

\bibitem{1}
M.~Alzenad and H.~Yanikomeroglu, ``Coverage and rate analysis for unmanned
  aerial vehicle base stations with {LoS/NLoS} propagation,'' \emph{2018 IEEE
  Globecom Workshops (GC Wkshps)}, pp. 1--7, 03 2018.

\end{thebibliography}
\bibliographystyle{IEEEtran}

\end{document}